\documentclass[draft]{agujournal2019}
\usepackage{url} 
\usepackage{lineno}
\usepackage[inline]{trackchanges} 
\usepackage{amsmath}
\usepackage{soul}
\usepackage{hhline}
\usepackage{tabularray}
\usepackage{multirow}
\draftfalse

\journalname{Journal of Advances in Modeling Earth Systems}

\begin{document}

%
%


\title{A Model Intercomparison Study of Mixed-Phase Clouds in a Laboratory Chamber}

%
%




\authors{Aaron Wang\affil{1}, 
Sisi Chen\affil{2}, 
Steve Krueger\affil{3}, 
Piotr Dziekan\affil{4}, 
Kotaro Enokido\affil{5}, 
Fabian Hoffmann\affil{6}, 
Agnieszka Makulska\affil{4}, 
Bernhard Mehlig\affil{7}, 
Gaetano Sardina\affil{8}, 
Grigory Sarnitsky\affil{7,8}, 
Silvio Schmalfuß\affil{9}, 
Shin-ichiro Shima\affil{5}, 
Fan Yang\affil{10}, 
Mikhail Ovchinnikov\affil{1}, 
Raymond A. Shaw\affil{11}}

\affiliation{1}{Pacific Northwest National Laboratory, Richland, Washington, USA}
\affiliation{2}{National Center for Atmospheric Research, Boulder, Colorado, USA}
\affiliation{3}{The University of Utah, Salt Lake City, Utah, USA}
\affiliation{4}{University of Warsaw, Faculty of Physics, Institute of Geophysics, Warsaw, Poland}
\affiliation{5}{University of Hyogo, Kobe, Hyogo, Japan}
\affiliation{6}{Freie Universit\"at Berlin, Berlin, Germany}
\affiliation{7}{University of Gothenburg, Gothenburg, Sweden}
\affiliation{8}{Chalmers University of Technology, Gothenburg, Sweden}
\affiliation{9}{Leibniz Institute for Tropospheric Research, Leipzig, Germany}
\affiliation{10}{Brookhaven National Laboratory, Upton, NY, USA}
\affiliation{11}{Michigan Technological University, Houghton, Michigan, USA}





\correspondingauthors{Aaron Wang}{aaron.wang@pnnl.gov}{Sisi Chen}{sisichen@ucar.edu}{Steve Krueger}{steven.krueger@utah.edu}



\begin{keypoints}
\item All models exhibit similar trends in ice microphysical properties with main discrepancies arising in the liquid phase.
\item Most models do not reach full glaciation except for those assuming a more well-mixed domain or simulating only the core region.
\item Compared to bin microphysics, Lagrangian schemes results in greater exposure of particles to supersaturation variability.

\end{keypoints}

%
%

%
%


\begin{abstract}
Mixed-phase clouds, composed of supercooled liquid droplets and ice crystals, play a critical role in weather and climate systems. Their complex microphysical interactions and coupling with turbulence at microscales govern the cloud properties at macroscales, yet remain challenging to observe and quantify under atmospheric conditions. This model intercomparison study utilizes ten model configurations to simulate mixed-phase cloud evolution in the Michigan Technological University's Pi Chamber. The models span a range of frameworks, including box models, direct numerical simulation, and large-eddy simulation models, and incorporate both bin and Lagrangian microphysics. Each model was tuned to reproduce the observed liquid-phase steady state prior to ice injection. Ice particles were then introduced into the domain at various rates to examine cloud glaciation behavior. By the intercomparison design, all models successfully reproduced the observed mean droplet radius and number concentration during the liquid-phase stage. Increasing ice particle injection rates led to consistent qualitative trends across models: depletion of liquid water, reduced total water content, and a shift in particle size distributions toward larger radii. However, quantitative differences arose due to variations in model treatment in dynamics and microphysics, including subgrid-scale turbulence parameterizations, wall forcing, and particle removal parameterizations. Most models that simulate the full chamber retained liquid droplets near the lower boundary, where supersaturation forcing is strongest and droplets are replenished before mixing into the core region. These surviving liquids droplets were absent in simulations assuming a well-mixed domain, excluding the near-wall region, or using coarse grid spacing.
\end{abstract}

\section*{Plain Language Summary}

Mixed-phase clouds are clouds that contain both liquid droplets and ice crystals. Their understanding is important for weather and climate prediction. This glaciation process depends on the interaction between droplets, ice crystals and water vapor, all mixed by turbulent motion of air. To study how clouds behave when the number of ice particles increases, we used ten different computer models to simulate experiments conducted in the Pi Cloud Chamber, a controlled laboratory environment. First, we adjusted each model to reproduce the observed cloud made only of liquid droplets. Then, we added ice particles at at different rates to observe the cloud glaciation process. All models showed similar trends, such as decrease of liquid water amount and the growth of ice particles. However, the models differed in details depending on how they represented turbulence, chamber walls, and removal of particles. Most models did not show complete glaciation, except for those with coarser grid resolutions or excluding near-wall regions rich in supercooled liquid. This study highlights that accurately simulating mixed-phase clouds requires capturing small-scale processes. Laboratory experiments like those in the Pi Cloud Chamber help scientists test and improve cloud models, which can eventually improve weather and climate predictions.


%
%

\section{Introduction}

Mixed-phase clouds, characterized by the coexistence of supercooled liquid and ice, play a crucial role in Earth's weather and climate systems \cite{Morrison2011,Villanueva2022,Korolev2017MM}. Their microphysical processes control phase transitions, precipitation efficiency, radiative properties, and cloud lifetimes, but modeling these processes remains challenging due to uncertainties in parameterizations used in coarse-resolution models \cite{Fan2011,Korolev2017MM,Korolev2022}. Atmospheric measurements are limited by transient cloud behavior and uncertain boundary conditions, hindering direct validation of model physics.

Controlled cloud chamber experiments, such as those conducted at Michigan Technological University's Pi Convection Cloud Chamber \cite{Chang2016BAMS,Desai2019GRL}, provide robust environments to study mixed-phase clouds under reproducible conditions. These facilities enable detailed measurements of aerosol, thermodynamic, and microphysical properties in statistically steady states. Combined with high-resolution numerical models like direct numerical simulation (DNS) and large-eddy simulation (LES), which can resolve fine-scale cloud processes such as supersaturation fluctuations and phase transitions, these tools offer critical insights for model development and validation.
Recognizing the need for such validation, a pioneering model intercomparison study using warm-phase cloud chamber experiments at the Pi Chamber was conducted by \cite{chen2025} at the 10th International Cloud Modeling Workshop \cite{xue2022}. This study evaluated a range of models, from DNS to LES and 1D turbulence models, to test their ability to simulate steady-state warm clouds under varying aerosol injection rates. While all models captured qualitative microphysical responses, significant quantitative discrepancies were found, such as differences in particle number concentrations due to variations in aerosol activation, supersaturation fluctuations, and particle removal mechanisms like sedimentation and diffusion. These findings underscored the importance of laboratory experiments in constraining models and highlighted the need for targeted studies to address specific physical processes.

Building on previous warm-phase research by \citeA{chen2025}, this study shifts focus to the complex dynamics of mixed-phase clouds in the Pi Chamber, exploring their transition from supercooled liquid to mixed-phase or fully glaciated states. Mixed-phase processes like ice nucleation, depositional growth, and the Wegener-Bergeron-Findeisen (WBF) process involve significant uncertainties, even in advanced atmospheric models. Experiments by \citeA{Desai2019GRL} demonstrated that steady-state mixed-phase clouds can be simulated in controlled environments, showing that the ice fraction is governed by injected ice-nucleating particle (INP) concentrations. Increased INP rates caused particle size distributions to evolve from single-mode liquid droplets to bimodal distributions, with decreasing mean droplet sizes and increasing ice fractions due to active WBF processes. Even at high INP rates, supercooled liquid persisted, with maximum ice mass fractions reaching 82\%.

Numerical studies have further demonstrated that spatial heterogeneity and turbulent fluctuations significantly impact glaciation processes \cite{Wang2024ACP}. For example, under identical ice injection rates, a bulk model reached full glaciation, while LES retained supercooled liquid, highlighting the importance of resolving turbulence-induced variability.

This study aims to evaluate how models represent mixed-phase cloud microphysics in controlled environments, with the goal of improving their accuracy in simulating turbulent mixed-phase clouds. Section 2 introduces the case setup and models, Section 3 presents the intercomparison results, and Section 4 discusses conclusions and implications for atmospheric modeling.

\section{Methods}
\label{sec:Method}
\subsection{Case Overview}
\label{sec:Overview}
This model intercomparison case is based on mixed-phase cloud experiments conducted in the Pi Chamber, following the setups described by \citeA{Desai2019GRL} and \citeA{Wang2024ACP}. The wall temperatures are $4^\circ\mathrm{C}$, $-12^\circ\mathrm{C}$, and $-16^\circ\mathrm{C}$ for the bottom, side, and top walls, respectively, resulting in a mean temperature below the freezing point. The Pi chamber can be configured as in either cylinder or rectangular geometries and is capable of generating and sustaining cloud formation under controlled conditions. The measurements were conducted in a cylindrical domain, but most models (including those used in this work) utilize a rectangular geometry. Measurements from the Pi Chamber experiments, including the top, bottom, and side wall temperature, as well as droplet mean radius and concentration at steady-state, were used to constrain the models' initial and boundary conditions. Model performance was evaluated using particle size distributions and domain-mean cloud properties, such as total water content and mean particle radius. Further details of the case description are provided in \citeA{chen_2024_15626802}.

\subsection{Procedures}
\label{sec:Procedures}
The numerical experiments follow three stages: a dry dynamic spin-up stage, a liquid-phase stage, and a mixed-phase stage. The spin-up stage aims to establish steady-state turbulence and thermodynamic conditions before aerosol injection. In the liquid-phase stage, sea salt aerosols with a diameter of 125 nm are used as cloud condensation nuclei (CCN) and are injected to form supercooled liquid droplets at sub-freezing temperatures. A steady-state in liquid-phase microphysics, characterized by a mean droplet radius of 7.75 $\mu$m and a number concentration of 25 cm$^{-3}$ is achieved by tuning the CCN injection rate. 
This stage is used for constraining and calibrating the liquid-phase condition before ice injection in mixed-phase stage. For models simulating the full chamber domain, both side-wall wetness and CCN injection rates are adjusted to match the observed droplet properties. For models simulating only the core region, a forcing on mean supersaturation is prescribed to represent the water vapor and heat exchange between the core and the surrounding area. In these core-only models particle sedimentation is implemented as a probabilistic removal due to the absence of chamber walls. Both the mean supersaturation forcing and particle sedimentation are tuned to obtain the observed droplet radius and number concentration. The CCN injection rate will remain constant throughout the entire simulation. 

After the liquid-phase steady state is established, ice crystals  with an initial radius of 2~$\mu$m are uniformly injected into the domain ~\cite{Wang2024ACP}. Ice particles, rather than INP \cite{Desai2019GRL}, are used in the model in order to eliminate uncertainties associated with ice nucleation and focus on ice growth process. Ice-particle injection rates range from 0.5 to 15 cm$^{-3}$ min$^{-1}$. In each simulation, ice is injected with a fixed rate until a new steady state is reached. The ice particles in the Pi Chamber are small enough to be reasonably assumed spherical in shape \cite{Desai2019GRL,Wang2024ACP}. Also because of the short particle lifetime in the Pi Chamber, collision and habit prediction are not considered in this study.

In \citeA{Desai2019GRL}, the particle size distributions were measured using a digital holographic instrument \cite<Holo-Pi,>[]{Desai2018JAS}. Holo-Pi is capable of determining both particle size and shape, the latter being useful for distinguishing liquid droplets from ice. Due to its pixel resolution of 3.5 $\mu$m, a theoretical resolution limit of 7 $\mu$m (corresponding to two pixels) is resulted.  Therefore, a cutoff radius of 3.5 $\mu$m was used for calculating the droplet and ice mean radii, number concentration, and water content. In practice, rejection of noise and identification of particles becomes more robust with increasing particle size up to approximately 6 $\mu$m, so the analysis in Section~\ref{sec:result_obs} used a higher cutoff radius.  
\citeA{Wang2024ACP} show that applying a cutoff radius affects the calculated mean droplet number concentration and radius, but the contribution of the sub-cutoff droplets to total liquid water content is small. Ice properties are barely affected because ice crystals generally exceed the cutoff radius due to their efficient growth. 


\subsection{Models}
\label{sec:Models}

Seven models were used in the intercomparison, including four LES models, one DNS model, one statistical model, and one bulk scalar mixing model (hereafter referred to as the ``bulk model''). The SAM-LES model was run with three different microphysics schemes, and the SCALE-SDM LES model was run with two different grid spacings, resulting in a total of ten sets of simulations. All LES models simulate the entire chamber with a rectangular geometry of 2 m $\times$ 2 m in the horizontal and 1 m in the vertical direction. The DNS and statistical models focus on the central, well-mixed region of the chamber, employing a cubic domain of 20 cm $\times$ 20 cm $\times$ 20 cm. The model configurations are summarized in Table~\ref{tab:models}, with detailed information provided below.

\begin{table}[]
\caption{Names and configurations of the model members.}
\includegraphics[width=\textwidth]{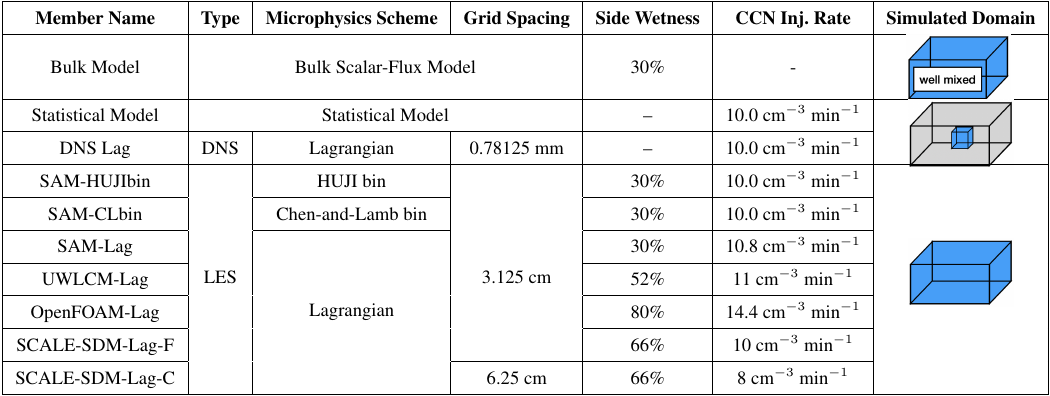}
\label{tab:models}
\end{table}

\textbf{The bulk model}, described in \citeA{Wang2024ACP}, solves budget equations for two scalars, temperature and water vapor mixing ratio, with an assumption of well-mixed (uniform) domain. This assumption is generally valid for Rayleigh-B\'enard convection, except near the chamber walls.
The mean supersaturation is computed as a function of the domain-average temperature, $\overline{T}$, and water vapor mixing ratio, $\overline{q}$.
The equilibrium values of $\overline{T}$ and $\overline{q}$ are determined by the net fluxes of sensible and latent heat from the chamber walls and droplets. A bulk aerodynamic formulation is used to model the area-integrated fluxes of temperature and water vapor from the walls. The flux velocity scale for the bottom and top walls, $V_0$, is derived from the vertical flux of sensible and latent heat implied by the Nusselt number. For the side walls, the flux velocity scale is assumed to be $0.42\, V_0$ and the side-wall wetness is 0.30
\cite{Wang2024ACP}. The net vapor flux to each type of particle (i.e., the net condensation or deposition rate) is proportional to the mean supersaturation with respect to the phase of the particle, $\overline{s}_p$, the particle number concentration, $N_p$, and the mean radius, $\overline{r}_p$,  which itself depends on $\overline{s}_p$ only \cite{Krueger2020ACP}.
The subscript $p$ indicates the phase of the particle (liquid or ice).
Given number concentrations and densities of droplets and ice, and the area and temperature of each wall, the bulk model predicts equilibrium values of $\overline{T}$ and $\overline{q}$, and therefore, of $\overline{s}_\text{ice}$, $\overline{s}_\text{liquid}$, $\overline{r}_\text{ice}$, and $\overline{r}_\text{liquid}$. Instead of injecting ice particles, we keep $\overline{N}_\text{liquid}$ as 25 cm$^{-3}$ and vary $\overline{N}_\text{ice}$, as conducted by \citeA{Wang2024ACP}.

\textbf{The statistical model}, described in detail by \citeA{sarnitsky2025does}, was developed for homogeneous isotropic turbulence and is therefore applied to the chamber's well-mixed core region. It uses a stochastic differential equation to simulate supersaturation fluctuations experienced by droplets and ice particles. This equation is derived from the mapping-closure mixing model~\cite{pope1991mapping}, assuming Gaussian supersaturation fluctuations in a well-mixed condition, where droplets and ice particles experience supersaturation fluctuations with identical statistics. DNS simulations of the same core region confirms both Gaussianity and the well-mixed assumption. A great challenge in formulating the model is to correctly represent the mixing between the core region and the rest of the chamber. To simulate this, both mean and variance of supersaturation are forced. The mean supersaturation for liquid, $\overline{s}_\text{liquid}$, is forced by adding a term $-(\overline{s}_\text{liquid}-\overline{s}_0)/\tau$ to the evolution equation for $\overline{s}$ \cite{Desai2019GRL, saito19}. Here $\tau=60\,{\rm s}$ is a relaxation time consistent with the time it takes for the chamber to relax to equilibrium \cite{chen_2024_15626802, Desai2019GRL}. The parameter $\overline{s}_0$ controls the target mean supersaturation, which is also affected by phase changes. In this study, $\overline{s}_0$ is set to 3\% to yield the prescribed steady-state droplet radii and number concentration in the liquid-phase stage, which is similar to the value (3.1\%) used by \citeA{Desai2019GRL}, and lower than the 5.97\% found by SAM LES \cite{Wang2024ACP} and suggested by the case description \cite{chen_2024_15626802}. However, 5.97\% is the prescribed steady-state value of the chamber's core region in the spin-up stage (before injecting CCN). During the liquid-phase stage, the mean supersaturation over the entire chamber decreases. Although $\overline{s}_0$ is time-dependent in reality, a constant value is used here for simplicity. 
The supersaturation variance is held constant at 2\%, as prescribed in the case description \cite{chen_2024_15626802}. To represent particle sedimentation from the core region to the rest of the chamber, particles are removed probabilistically based on their Stokes settling speeds \cite{chen_2024_15626802}. It is found that reducing the settling speed by a factor of $\alpha=0.8$ compared with the case description improves agreement on target droplet radii and numbers in the liquid-phase stage. The CCN injection rate is taken to be $10$ cm$^{-3}$ min$^{-1}$, as specified in the case description. 
Note that the previously published statistical model simulation of the Pi chamber by \cite{sarnitsky2025does} used $\overline{s}_0 = 5.97\%$ and $\alpha = 1.0$, and the linearization of the supersaturation function was done near $\overline{s}_0 = 5.97\%$, whereas the current study uses 0\%.

\textbf{The DNS} with Lagrangian droplet and ice particles (hereafter referred to as DNS-Lag) is also designed to model homogeneous isotropic turbulence, thus is applied to simulate the chamber's core region. A detailed description of the model is provided by \citeA{sarnitsky2025does}. DNS-Lag resolves a single supersaturation field governed by a convection-diffusion-reaction equation, rather than two separate fields for water mixing ratio and temperature. This simplification requires supersaturation fluctuations to be small, allowing linearization of the supersaturation function with respect to water vapor mixing ratio. This assumption holds in the well-mixed core region. Additionally, the single supersaturation approach assumes equal diffusivities for temperature and water vapor, which is not an accurate approximation near the walls of the chamber or for the study of the inhomogeneous flow on the scale of the whole chamber \cite{Chandrakar2020JFM}. However, in the core region away from the strong near-wall gradients, this assumption is justified (e.g., Fig.~8 in \cite{Chandrakar2020JFM}), provided the correct supersaturation variance is supplied, as prescribed in the case description \cite{chen_2024_15626802}. And the accuracy of the single-supersaturation approach has been validated through comparison with the two-equation approach in previous DNS studies ~\cite{lanotte2009cloud,sardina2015continuous, sarnitsky2025does}. DNS-Lag uses a standard pseudospectral solver \cite{sardina2015continuous, fries2021key} with triply periodic boundary conditions. To represent mixing between the core region and the rest of the chamber, supersaturation forcing is applied similar to the statistical model. Droplets and ice particles are advected as tracers by the velocity field. Particle removal due to sedimentation is modeled probabilistically using the same approach and parameters as in the statistical model. 

All simulations from the statistical model and DNS-lag, including the one to establish the liquid-phase steady state, were conducted for 10 minutes. A  30-seconds spinup period was used to establish the fully developed turbulence in DNS-Lag prior to the CCN injection in liquid-phase stage.
 
\textbf{The SAM model} \cite<System for Atmospheric Modeling, >[]{Khairoutdinov2003JAS}, with the single-moment bin microphysics scheme developed by the Hebrew University of Jerusalem group \cite{Khain2004JAS}, was introduced and applied in \citeA{Wang2024ACP}. This model is referred to as SAM-HUbin hereafter. SAM has been widely used to simulate cloud chambers and their wall fluxes \cite{Thomas2019JAMES,Yang2022JAMES,Yang2023JAMES,Yang2025ACP,Wang2024ACP,Wang2024Entrainment,Wang2024JAMES,Wang2024JAS,Wang2025PoF,Wang2025JAMES,Thomas2025JAMES}. The velocities are solved on an Arakawa staggered C-grid \cite{Arakawa1977GCMA}, advected with a second-order central scheme, and dissipated with a turbulent kinetic energy (TKE) subgrid-scale (SGS) model \cite{Deardorff1980BLM}. The scalars are advected by a multidimensional positive definite advection transport algorithm \cite{Smolarkiewicz1990JCP} and diffused with a turbulent Prandtl number of 1, following the Reynolds analogy \cite{Kays1980}. The bin microphysics includes 33 mass-doubling bins for CCN, another 33 for liquid, and another 33 for ice. The only difference between the current study and the SAM-HUbin simulations in \citeA{Wang2024ACP} is the simulation procedure: \citeA{Wang2024ACP} performs a single long simulation with varying ice injection rates, whereas this study performs seven independent simulations, each with a different ice injection rate. 
Each simulation consists of three 20-minute stages: (1) spin up of moist turbulence, (2) liquid-phase stage with CCN injection, and (3) mixed-phase stage with injection of ice. 
To match the observed droplet radius and number concentration, the side-wall wetness with respect to ice is set to 0.30, and the CCN injection rate is 10 cm$^{-3}$ min$^{-1}$. 

The second set of SAM LES was conducted with a different bin microphysics scheme by \citeA{chen1994simulation}. This scheme has been implemented in SAM for Pi chamber simulations by \citeA{Yang2022JAMES,Yang2023JAMES,Yang2025ACP}. This model is referred to as SAM-CLbin. To maintain consistency with SAM-Hubin, SAM-CLbin also uses 33 mass-doubling bins of dry CCN, droplets, and ice. Unlike SAM-HUbin, which used a single moment scheme (tracking only mass), SAM-CLbin treats both number and mass concentrations as prognostic variables in each bin. CCN activation follows a Twomey-type parameterization, in which CCN become droplets when the environment supersaturation exceeds their critical supersaturation \cite{Yang2023JAMES}. The rest of the model setups, including CCN injection rates and side-wall wetness, are identical to those of the SAM-HUbin.

The third set of SAM LES uses a Lagrangian particle method, referred to as SAM-Lag. It shares the same dynamical core as the other two SAM LES models described above. The Lagrangian cloud microphysical scheme originates from \citeA{hoffmann2015} and has been progressively refined and applied in subsequent studies \cite<e.g.,>[]{hoffmann2019,hoffmann2020}. SAM-Lag was recently compared to SAM-CLbin in a warm-phase cloud chamber setup \cite{Yang2023JAMES}. Similar to other Lagrangian cloud microphysical schemes, this scheme uses individually simulated computational particles, each representing a multitude of identical hydrometeors. Cloud microphysical process rates (condensation/evaporation, deposition/sublimation) are scaled accordingly to represent the corresponding effect on the thermodynamic fields of SAM, with which the cloud microphysics are two-way coupled to the SAM dynamical core. For the transport of the computational particles, velocity fields are interpolated to particle positions using a scheme that maintains the incompressibility of the flow \cite<cf.>[]{grabowski2018}. The particle growth equation for haze and droplets considers both curvature and solute effects \cite{hoffmann2015}. For this intercomparison several adaptations were made in the cloud microphysical scheme: ice crystals are assumed to be spherical for calculating deposition and sublimation; nucleation processes are excluded, with ice crystals injected with a prescribed radius of $2\,\mu\text{m}$; and particle sedimentation is modeled using a simple velocity parameterization proportional to the square of particle radius for all hydrometeors. The number of computational particles is not fixed, but a balance of particle injections and sedimentation. During liquid-phase stage (CCN injection), about $25$ computational particles per grid box represent hydrometeors. In mixed-phase stage (with ice injection), the number of computational particles per grid box increases to about $30$, $38$, $51$, $75$, $110$, and $120$ for ice injection rates of $0.5$, $1.5$, $3.0$, $5.0$, $10.0$, and $15\,\text{cm}^{-3}\,\text{min}^{-1}$, respectively. Note that the concentration of computational particles did not converge for injection rates $\ge 5.0\,\text{cm}^{-3}\,\text{min}^{-1}$. To match the prescribed droplet size and concentration, SAM-Lag uses a side-wall wetness of $0.3$ and a CCN injection rate of $10.8\,\text{cm}^{-3}\,\text{min}^{-1}$. 
To gain reliable statistics, each simulation includes a 5\,\text{min} spin-up period without hydrometeor, followed by $40\,\text{min}$ of liquid-phase stage (CCN injection), followed by $20\,\text{min}$ of mixed-phase stage (CCN and ice injection).

\textbf{The UWLCM model} \cite<University of Warsaw Lagrangian Cloud Model, >[]{Dziekan2019} is an LES model with an Eulerian dynamical core and Lagrangian particle microphysics, similar to SAM. This LES model is referred to as UWLCM-Lag. Advection is solved with the multidimensional positive-definite advection transport algorithm \cite{Smolarkiewicz1990JCP} on a dual, Arakawa C-grid. Smagorinsky model is used for SGS turbulence \cite{Smagorinsky1963}. Droplets, ice crystals and CCN are modeled using the particle-based scheme by \citeA{Arabas2015}, which implements the Super-Droplet Method (SDM) \cite{Shima2009QJRMS}. Each simulation begins with a 5-min spin-up period, followed by a 12-min of liquid-phase period to reach steady-state droplet statstics, and a 10-minute mixed-phase period. To match the observed droplet radius and number concentration, sidewall wetness is set to 0.52, and the CCN injection rate is 11~cm$^{-3}$min$^{-1}$. Particles are injected uniformly in each grid cell. Each super-droplet represents 2 real liquid droplets or CCN, which results in approximately 450 super-droplets per grid cell in liquid-phase stage. In the mixed-phase stage, spherical ice super-particle representing 1 or 2 real ice crystals are injected. The total number of computational particles per grid cell during the mixed-phase stage ranges between $500$ and $1600$ depending on the ice injection rates. Rigid boundary conditions are applied, and particles are removed upon contact with the walls.

\textbf{The OpenFOAM} (Open Field Operation And Manipulation) with Lagrangian microphysics, referred to as OpenFOAM-Lag, is another LES model. A finite volume method solver named buoyantPimpleFoam from OpenFOAM's development branch (version 20200614) served as base for the simulation of the continuous phase, and was extended to include a transport equation for water vapor. The governing equations are solved with second order accurate schemes, and the Smagorinsky model was used for SGS turbulence \cite{Smagorinsky1963}. For particle tracking, OpenFOAM's built-in Lagrangian library, the kinematicCloud, was used and extended to include cloud microphysics and heat and mass transfer with the continuous phase, following the approach of  \citeA{niedermeier2020characterization}. Heat transfer is calculated using the correlation from \citeA{ranz1952evaporation}. The model includes deliquescence/efflorescence effects, and droplet growth follows Köhler theory. Ice particle growth follows the equation in \citeA{chen2023mixed}. Droplets and ice particles are injected randomly throughout the domain at every time step. OpenFOAM uses a parcel approach \cite<e.g.,>[]{gosman1983aspects}, similar to the SDM. Each parcel represents a multitude of real particles with identical properties: five for CCN and liquid droplets; and one to nine for ice crystals depending on the injection rates (1, 1, 2, 3, 6, 9 for injection rates of $0.5$, $1.5$, $3.0$, $5.0$, $10.0$, and $15\,\text{cm}^{-3}\,\text{min}^{-1}$, respectively). Particle motion is computed by integrating Newton's second law to obtain their velocity and from that their displacement per time step. Forces acting on the particles include the sphere drag force with an empiric drag coefficient correlation from \citeA{schiller1933uber} and a correction for small particles according to \citeA{cunningham1910velocity}, the gravitational force, and a force representing Brownian motion. Both the spin-up stage and liquid-phase stage are 20 minutes each, the mixed-phase stage is run for ten minutes. As in UWLCM, particles are removed upon contact with walls.

\textbf{The SCALE-SDM}, referred to as SCALE-SDM-Lag, is an LES with Lagrangian microphysics. It consists of two components: SCALE (Scalable Computing for Advanced Library and Environment) and SDM. SCALE is a versatile numerical model designed for weather and climate simulations on Earth and other planets \cite{Nishizawa2015,Sato2015}. SCALE solves the compressible Navier–Stokes equations for moist air using a finite volume method on an Arakawa-C staggered grid \cite{Arakawa1977}. In the current setup, advection of dynamical variables is calculated with a fourth-order central difference scheme, while tracers are advected using the third-order upwind scheme with Koren's filter \citeyear{Koren1993}. A second-order central difference scheme is employed for other spatial derivatives. Time integration is performed using the fourth order Runge–Kutta scheme for dynamical variables and the three-step Runge–Kutta scheme by \citeA{Wicker2002} for tracers. The SGS turbulence parameterization employs the Smagorinsky scheme \cite{Smagorinsky1963} with the corrections of \citeA{Brown1994} and \citeA{Scotti1993}. To ensure numerical stability, a fourth-order hyper-diffusion term is applied, with a nondimensional coefficient set to $10^{-4}$ \cite{Nishizawa2015}.
SCALE incorporates SDM for resolving CCN, cloud, and precipitation particle microphysics \cite{Shima2009QJRMS,Shima2020}. 
Condensation and evaporation are solved using an implicit Euler scheme to handle stiffness.
SCALE-SDM-Lag is tested with two grid spacings: 3.125 cm (fine resolution, referred to as SCALE-SDM-Lag-F) and 6.25 cm (coarse resolution, referred to as SCALE-SDM-Lag-C). 
Each simulation consists of three 30-min stages: a dynamical spin-up period, a liquid-phase stage, and a mixed-phase stage. 
For SCALE-SDM-Lag-F, the sidewall wetness with respect to ice is set as 0.66, and the CCN injection rate is 10~cm$^{-3}$ min$^{-1}$ to match the observed droplet radius and number concentration. The time steps for dynamics, tracer advection, and microphysics are $0.5\times10^{-4}$ s, 0.02 s, and 0.02 s, respectively. Both sea salt and ice super-droplets are injected every 0.02 s with a rate of 10 super-droplets per grid box per min.  
For SCALE-SDM-Lag-C, to match the observed droplet radius and number concentration, the sidewall wetness is set the same as in SCALE-SDM-Lag-F, with a slightly lower CCN injection rate (8~cm$^{-3}$min$^{-1}$). The time steps for dynamics, tracer advection, and microphysics are $1.0\times10^{-4}$ s, 0.04 s, and 0.04 s, respectively. Both CCN and ice super-droplets are injected every 0.04 s with the rate of 10 super-droplets per grid box per min.

\section{Results}

Section \ref{sec:result_obs} presents a comparison between observations from \citeA{Desai2019GRL} and the models performed in this work. Because observations have challenges to differentiate between ice and liquid, the comparison is based on the statistics of the total particle size distribution (including both liquid and ice) above a cutoff radius limited by the instrumentation's resolution (see Section \ref{sec:Procedures}). Section \ref{sec:result_model} analyzes modeled liquid and ice properties separately across the models to assess the glaciation behavior.

\begin{figure*}[t]
    \includegraphics[width=\textwidth]{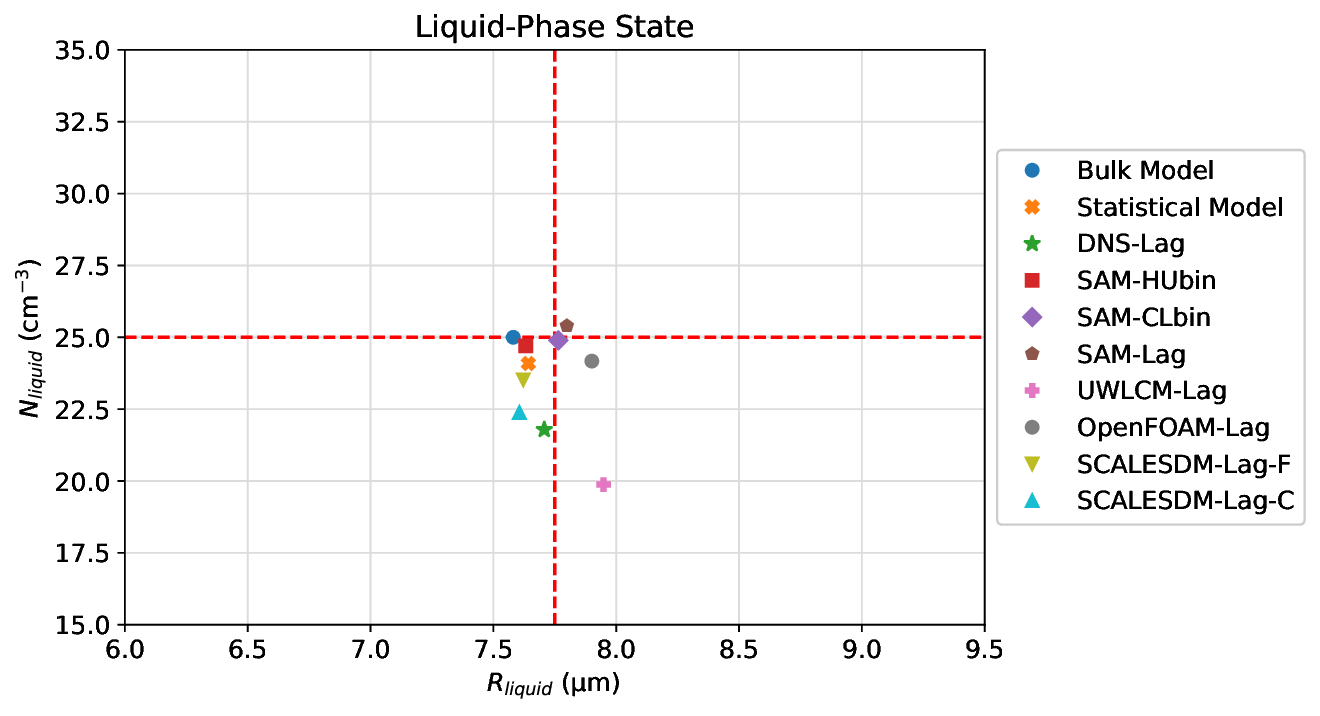}
    \caption{The mean droplet radius and number concentration at steady-state during the liquid-phase stage by each model compared to the Pi Chamber observation. The observational constraints on droplet radius ($7.75\ \mu m$) and number concentration ($25\ cm^{-3}$) are indicated by red dashed lines.}
    \label{fig:rl_NCL}
\end{figure*}

\subsection{Comparison of observation and models}
\label{sec:result_obs}

In the liquid-phase steady state (i.e., prior to ice injection), all models showed reasonable agreement with the observed mean droplet radius (7.75 µm) and number concentration (25 cm$^{-3}$) by tuning the model parameters detailed in Section \ref{sec:Models} (Fig.~\ref{fig:rl_NCL}). Note that the variation of droplet number concentration in the simulated domain can easily exceed 10 cm$^{-3}$ \cite<see Fig.~4 in>[]{Wang2024ACP}. This tuning is a necessary procedure to achieve a consistent liquid-phase starting point, given the inter-model variations in wall models and microphysics schemes. 

\begin{figure*}[t]
    \includegraphics[width=\textwidth]{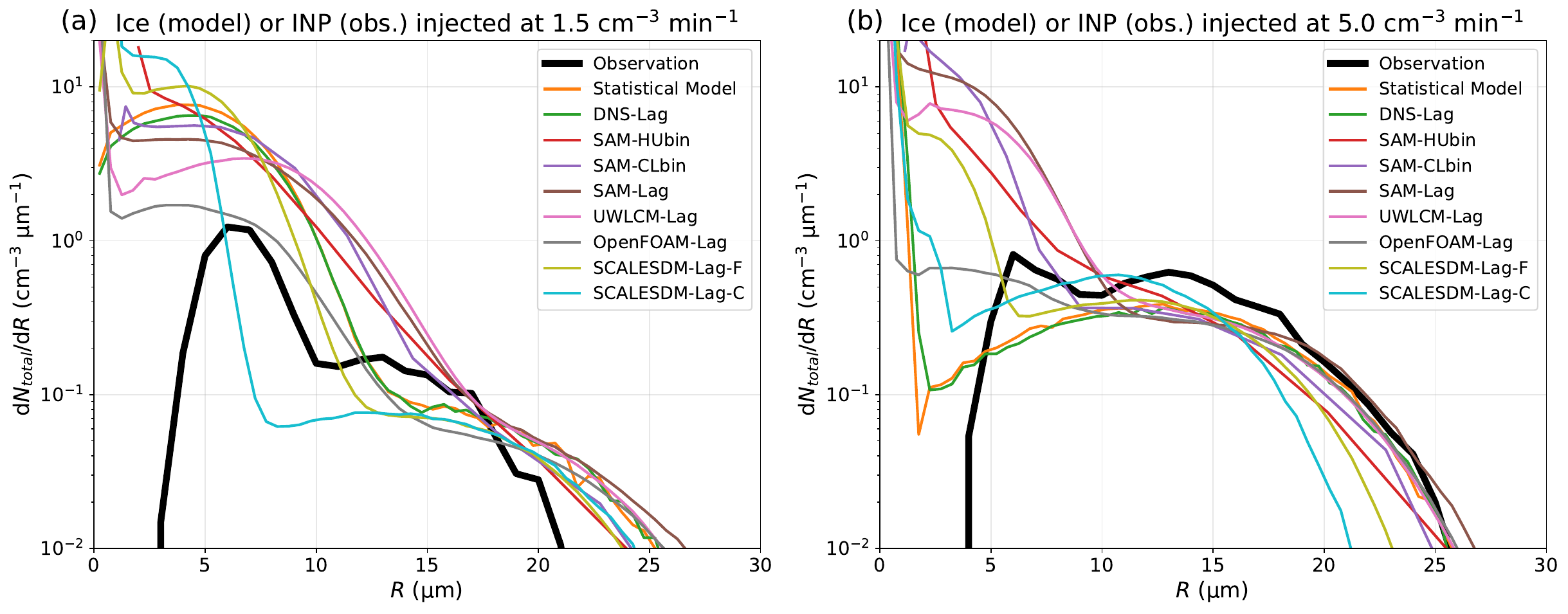}
    \caption{The total particle size distribution (including liquid and ice) at two injection rates of ice (for models) or INP (for observation).}
    \label{fig:psd_obs}
\end{figure*}

Fig. \ref{fig:psd_obs} shows particle size distributions under two INP (ice) injection rates in the observation (models). Both observation and models reveal bimodal distributions, with the first mode indicating droplets and the second mode ice. Both modes shift upward and leftward as the INP (ice) injection rate increases. In models that simulate only the chamber's core region (statistical model and DNS-Lag), the first mode in Fig.~\ref{fig:psd_obs}b is narrow and sharp, with the mode radius below the cut-off radius of 3.5 $\mu m$, implying full glaciation with only ice particles (second mode) left. In contrast, all models that simulate the entire chamber, except for SCALESDM-Lag-C, show a broader left mode, indicating a presence of droplets remaining in the domain. See also the discussion of~Fig.~\ref{fig:SS_Rli}.
Because not all INP are nucleated in the Pi Chamber, later we will compare the trend using the particle number concentration instead of the INP injection rate.

\begin{figure*}[t]
    \includegraphics[width=\textwidth]{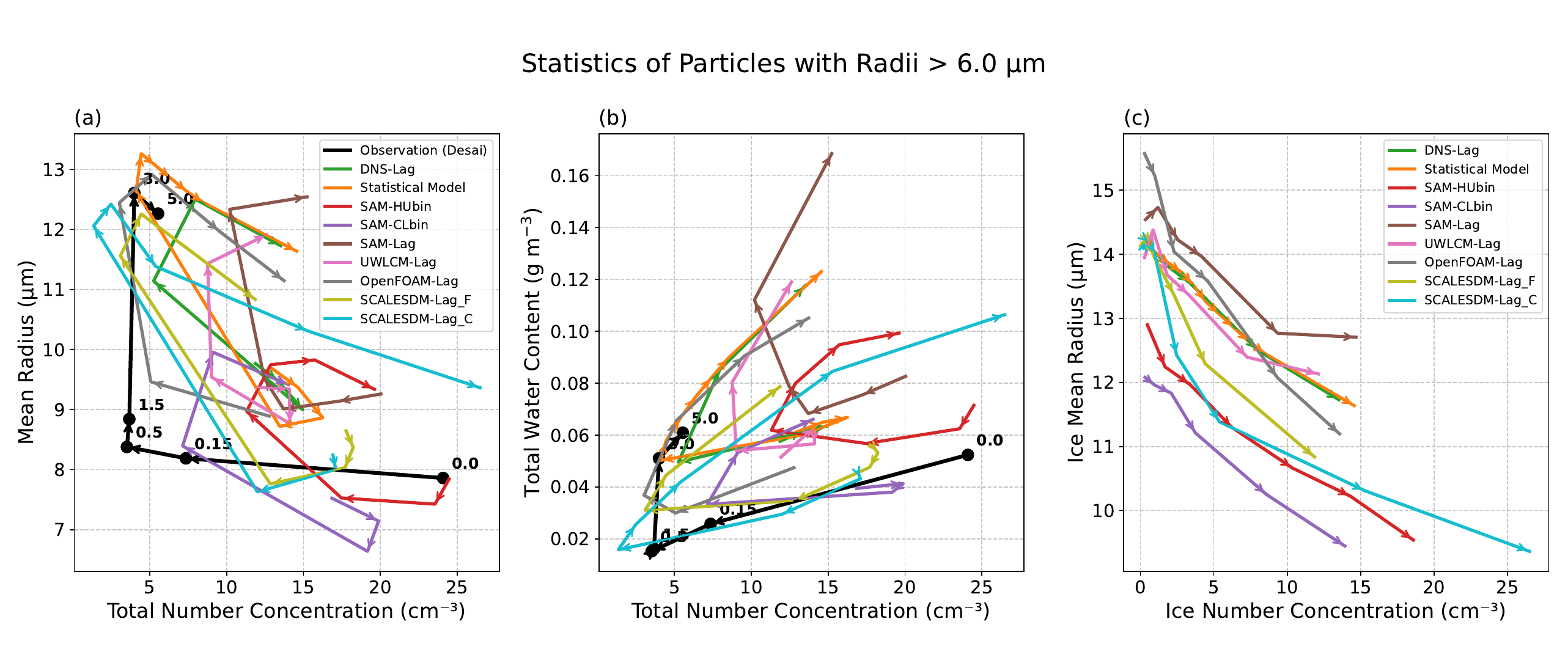}
    \caption{The relationships between (a) mean particle radius and number concentration, (b) total water content and number concentration, and (c) ice radius and ice number concentration (simulated results only). For total water content calculation, ice is assumed to have the same density as liquid water. Black curves with labeled values indicate observed INP injection rates (unit: cm$^{-3}$ min$^{-1}$), and arrows along each curve indicate the direction of increasing ice (model) or INP (observation) injection rate.}
    \label{fig:n_r_lwc_obs}
\end{figure*}

Additionally, the observed number concentration in the first mode (liquid) is significantly less than those in the models. As mentioned earlier, Holo-Pi has a roll-off in the detection probability for droplet radii below approximately 6 $\mu$m.  To ensure a reliable comparison, this higher cutoff radius should be used. The particle size distributions (PSDs) in \citeA{Desai2019GRL} indicate that all cases exhibit a decreased number concentration when the radius is less than 6~$\mu$m, consistent with this instrument limitation.

Accordingly, we compare the microphysical statistics from observation and models using only particles with  radii greater than 6~$\mu$m. For consistency, we assume that the ice crystals have the same density as liquid for calculating total water content, as holography cannot distinguish between ice and liquid. The results are presented in Fig.~\ref{fig:n_r_lwc_obs}, which reveals similar trend in both observation and models. Specifically, as the INP (ice) injection rate increases, the evaporation of liquid first causes a reduction of total water content and the total number concentration (Fig.~\ref{fig:n_r_lwc_obs}b). The mean radius increases due to an efficient ice growth (Fig.~\ref{fig:n_r_lwc_obs}a). At an intermediate injection rates (e.g, INP injections rates between 0.5-3.0 cm$^{-3}$ min$^{-1}$ in the observations), the decrease in droplet number is roughly balanced by the increased ice number, leading to a nearly constant total number concentration and an increase in mean radius (Fig.~\ref{fig:n_r_lwc_obs}a) and total water content (Fig.~\ref{fig:n_r_lwc_obs}b). At higher injection rates, the increase of crystal number exceeds the droplet loss, causing total number concentration and total water content to increases (Fig.~\ref{fig:n_r_lwc_obs}b), while the mean radius decreases owing to the reduced supersaturation as a results of increasingly strong competition among ice particles (Fig.~\ref{fig:n_r_lwc_obs}a). Figure~\ref{fig:n_r_lwc_obs}c confirms that ice radius decreases with increasing ice number concentration. Although the shapes of the curves vary among models, similar turning points are observed in both the mean radius and total water content, consistent with the observed trends.

\subsection{Analysis of liquid and ice in the models}
\label{sec:result_model}

\begin{figure*}[t]
    \centering
    \includegraphics[width=0.7\textwidth]{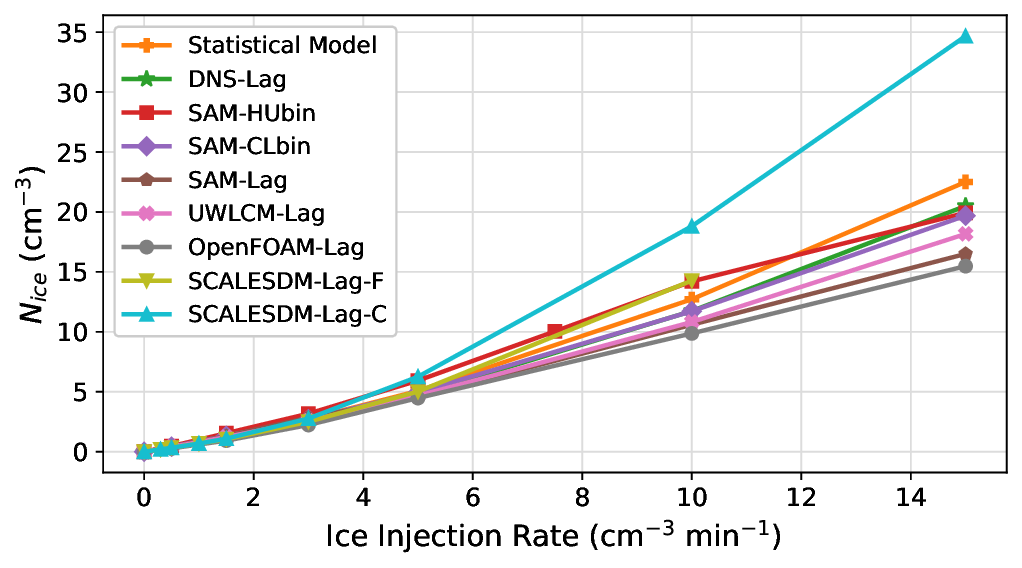}
    \caption{Ice number concentration at steady-state versus the ice injection rate given by each model.}
    \label{fig:IceIR}
\end{figure*}

As the same ice injection rate can produce different ice number concentrations across models, we use $N_\text{ice}$ as a more reliable metric to verify inter-model differences. Figure~\ref{fig:IceIR} shows the resulting $N_\text{ice}$ from all models except for the bulk model, which specifies the resulting $N_\text{ice}$ directly instead of injecting ice. All models agree on the trend of increasing $N_\text{ice}$ with the ice injection rate. SCALE-SDM-Lag-C with a coarser grid spacing apparently produces the highest $N_\text{ice}$. One possible explanation is that the coarse grid spacing results in enhanced mixing, thus causing faster glaciation.

\begin{figure*}[t]
    \includegraphics[width=\textwidth]{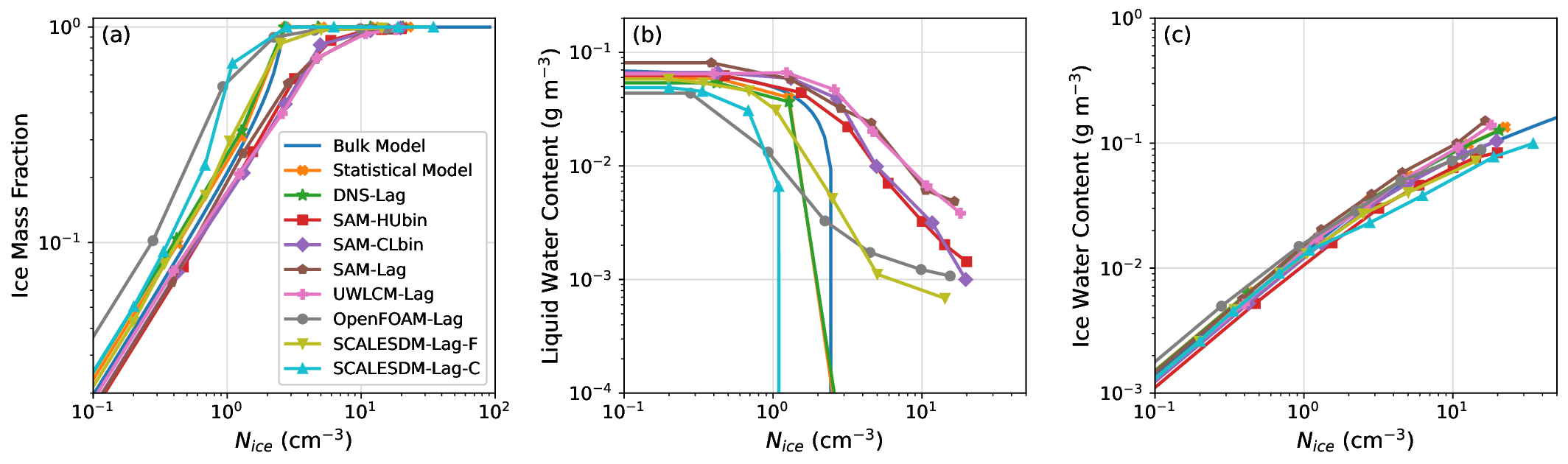}
    \caption{Ice mass fraction (a), Liquid water content (b), and ice water content (c) versus ice number concentrations.}
    \label{fig:IMF}
\end{figure*}

Change in ice mass fraction as a function of $N_\text{ice}$ is shown in Fig.~\ref{fig:IMF}a. Here, an ice mass fraction of 1 indicates full glaciation. Although all models follow similar trends, the bulk model reaches full glaciation at a distinct threshold, whereas the other models approach full glaciation more gradually as $N_\text{ice}$ increases. All models agree well in the ice water content (Fig.~\ref{fig:IMF}c), indicating that model differences in ice mass fraction are primarily due to variations in liquid-phase processes (Fig.~\ref{fig:IMF}b).

\begin{figure*}[t]
    \includegraphics[width=\textwidth]{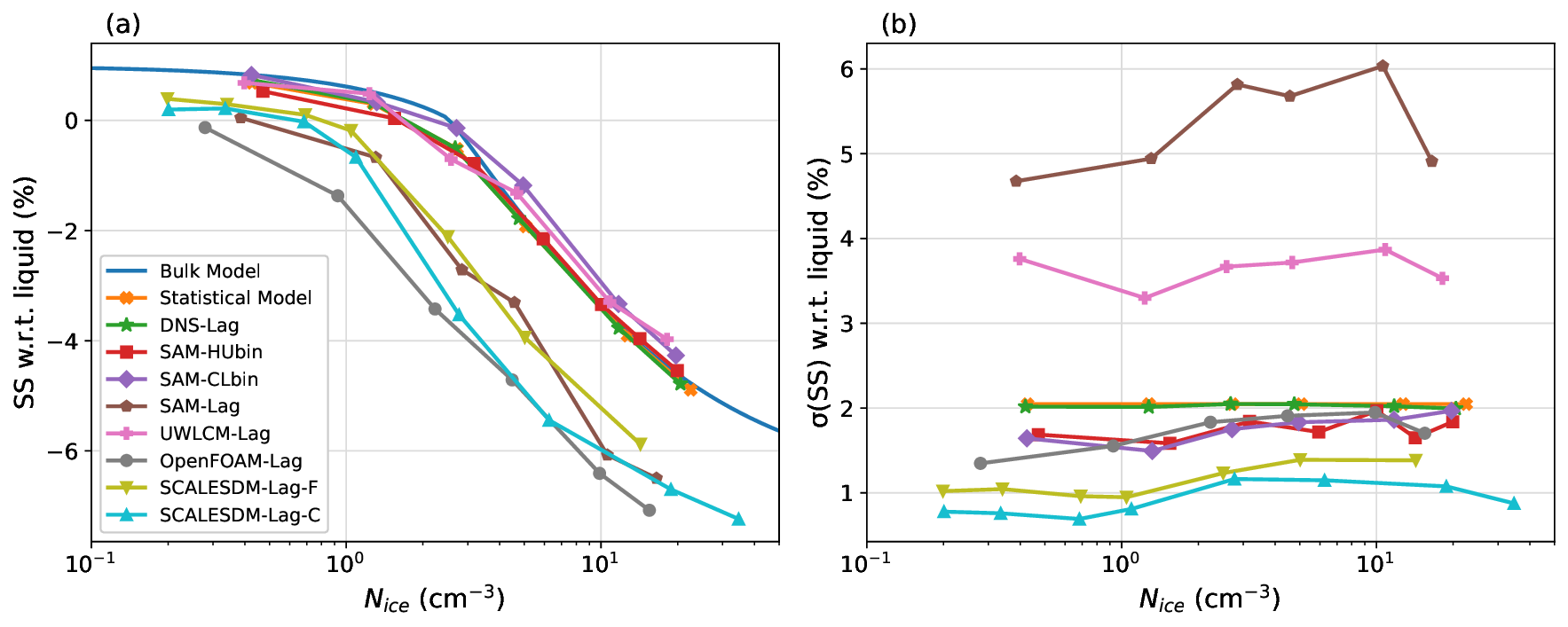}
    \caption{The (a) mean supersaturation and (b) standard deviation of supersaturation with respect to liquid versus ice number concentration in the bulk area excluding 12.5 cm near the walls.}
    \label{fig:SS}
\end{figure*}

Near the glaciation threshold, Fig.~\ref{fig:IMF}b--c illustrate that changes in slope is more pronounced in the liquid water content. Among the models, OpenFOAM consumes liquid water most efficiently, followed by SCALE-SDM and then the other models, which is consistent with the order of glaciation rates shown in Fig.~\ref{fig:IMF}a. In SCALE-SDM, coarser grid spacing results in more efficient glaciation due to enhanced mixing within a coarser grid scale. OpenFOAM also shows the lowest liquid water content (Fig.\ref{fig:IMF}b), which may be related to the supersaturation: the mean supersaturation in the bulk area in OpenFOAM is negative for all ice injections, whereas other models reach negative only when ice number concentrations exceed certain values (Fig.~\ref{fig:SS}a).

Cloud glaciation is strongly affected by supersaturation with respect to liquid. Because supersaturation has strong variations near the walls, for LES models that include the walls, the region within a distance of 12.5 cm from the walls is excluded from the calculation of mean and standard deviation of supersaturation. Fig.~\ref{fig:SS}a shows that mean supersaturation generally decreases with increasing $N_\text{ice}$, eventually falling below zero at certain points. To accurately represent mean supersaturation for models that simulate only core regions, it is crucial to account for the exchange of water vapor and temperature between the core region and the rest of the chamber. For the statistical model and DNS-Lag, this requires careful tuning the forcing scheme $\overline{s}_0$ (Section~\ref{sec:Models}).

In contrast to the mean supersaturation, supersaturation fluctuations show no clear trend with ice number concentration (Fig.~\ref{fig:SS}b). The bulk model assumes a well-mixed domain and does not consider supersaturation fluctuations. The statistical model and DNS-Lag use a prescribed constant supersaturation fluctuation. Among the LES models, SCALE-SDM exhibits the lowest supersaturation fluctuation, which reduces further with coarser grid spacing due to resolving less turbulence. SAM-Lag exhibits the highest supersaturation fluctuation, followed by UWLCM-Lag. SAM-HUbin, SAM-CLbin, and OpenFOAM-Lag show an intermediate level. Interestingly, OpenFOAM-Lag maintains a negative mean supersaturation for all ice injections, yet liquid water persists due to supersaturation fluctuations and strong forcing near the bottom (as shown later in Fig.~\ref{fig:SS_profile}a).

\begin{figure*}[t]
    \includegraphics[width=\textwidth]{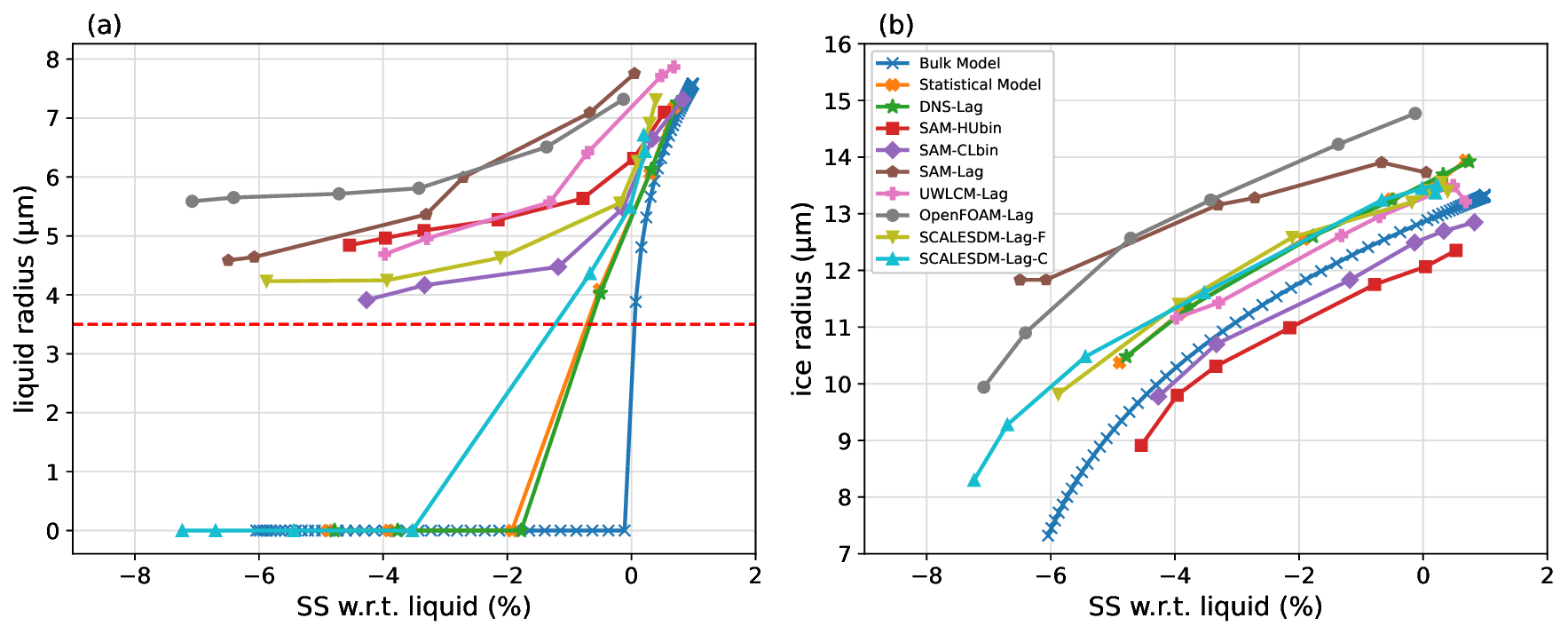}
    \caption{The mean (a) liquid and (b) ice radii versus supersaturation with respect to liquid. The red dashed line indicates the cut-off radius}
    \label{fig:SS_Rli}
\end{figure*}

The relationship between mean particle size and supersaturation is examined in Fig.~\ref{fig:SS_Rli}. Both liquid and ice particles decrease in size as supersaturation reduces. When the mean liquid radius falls below the cutoff radius of 3.5~$\mu$m, it indicates that no droplets above the detection threshold remain, signifying full glaciation. Figure~\ref{fig:SS_Rli}a shows that, in addition to the bulk model, which has previously been shown full glaciation \cite{Wang2024ACP}, DNS-Lag, the statistical model, and SCALE-SDM-Lag-C also predict full glaciated clouds. The reasons are as follows: As discussed in Section~2.2, DNS-Lag and the statistical model describe only the core region of the chamber. While the models account for particle sedimentation, they do not account for particle entrainment from regions outside the core. This is often acceptable, since the chamber is approximately well-mixed and particle properties in the core resembles those in the broader chamber domain, excluding near-wall regions. However, for the current case, the bottom of the chamber remains persistently supersaturated with respect to liquid (see Fig.~\ref{fig:SS_profile}). This allows continuous CCN activation. Droplets produced in this region can be advected to the core, sustaining liquid phase there. This is a mechanism not captured by core-only models. SCALE-SDM-Lag-C by contrast likely predicts early full glaciation due to an enhanced mixing within its coarser grid spacing, which homogenizes the domain. A comparison with SCALE-SDM-Lag-F reveals that refining the grid spacing increases both supersaturation and its fluctuations (Fig.~\ref{fig:SS}), as well as liquid size (Fig.~\ref{fig:SS_Rli}), thereby preventing full glaciation. Figure~\ref{fig:SS_Rli}b demonstrates that among the LES models, those using Lagrangian microphysics generally produce larger ice particles and lower number concentrations compared to bin microphysics. This maybe due to a weaker correlation between Lagrangian particles and the supersaturation field compared to bin microphysics (as further illustrated in Fig.~\ref{fig:SS_slices}), allowing particles to experience greater supersaturation variations and grow larger \cite{Chandrakar2016PNAS,MacMillan2022PRF}. For core-only models, the statistical model and DNS-Lag include a simplified particle removal scheme to represent sedimentation. A reduction in particle sedimentation rate compared to the Stokes removal specified in \citeA{chen_2024_15626802} is made to match the target liquid-phase steady state (Section~\ref{sec:Models}). This adjustment allows larger particles to remain longer in the simulation domain, increasing the mean ice radius. However, it does not affect the mean liquid radius, as liquid are smaller and less influenced by sedimentation.

\begin{figure*}[t]
    \includegraphics[width=\textwidth]{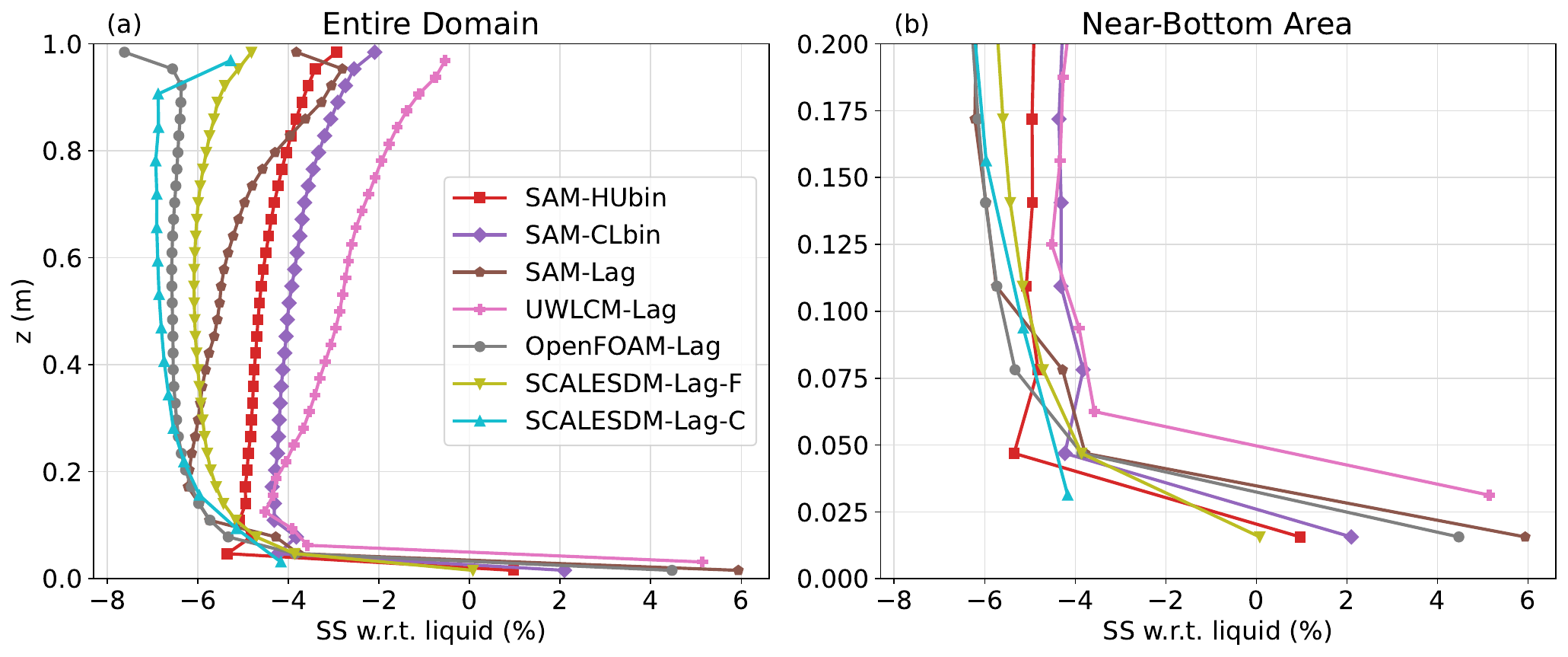}
    \caption{The vertical profile of supersaturation with respect to liquid as ice injection rate is 10 cm$^{-3}$ min$^{-1}$.}
    \label{fig:SS_profile}
\end{figure*}

Previous work by \citeA{Wang2024ACP} shows that liquid droplets tend to survive near the bottom region, where moisture and heat fluxes are strongest, creating a supersaturated area. Figure~\ref{fig:SS_profile} confirms that all LES models agree on this vertical structure, with supersaturation peaking near the bottom. However, SCALE-SDM-Lag-C, due to a coarser grid spacing, failed to resolve this localized supersaturation, resulting in complete glaciation.

\begin{figure*}[t]
    \centering
    \includegraphics[width=0.8\textwidth]{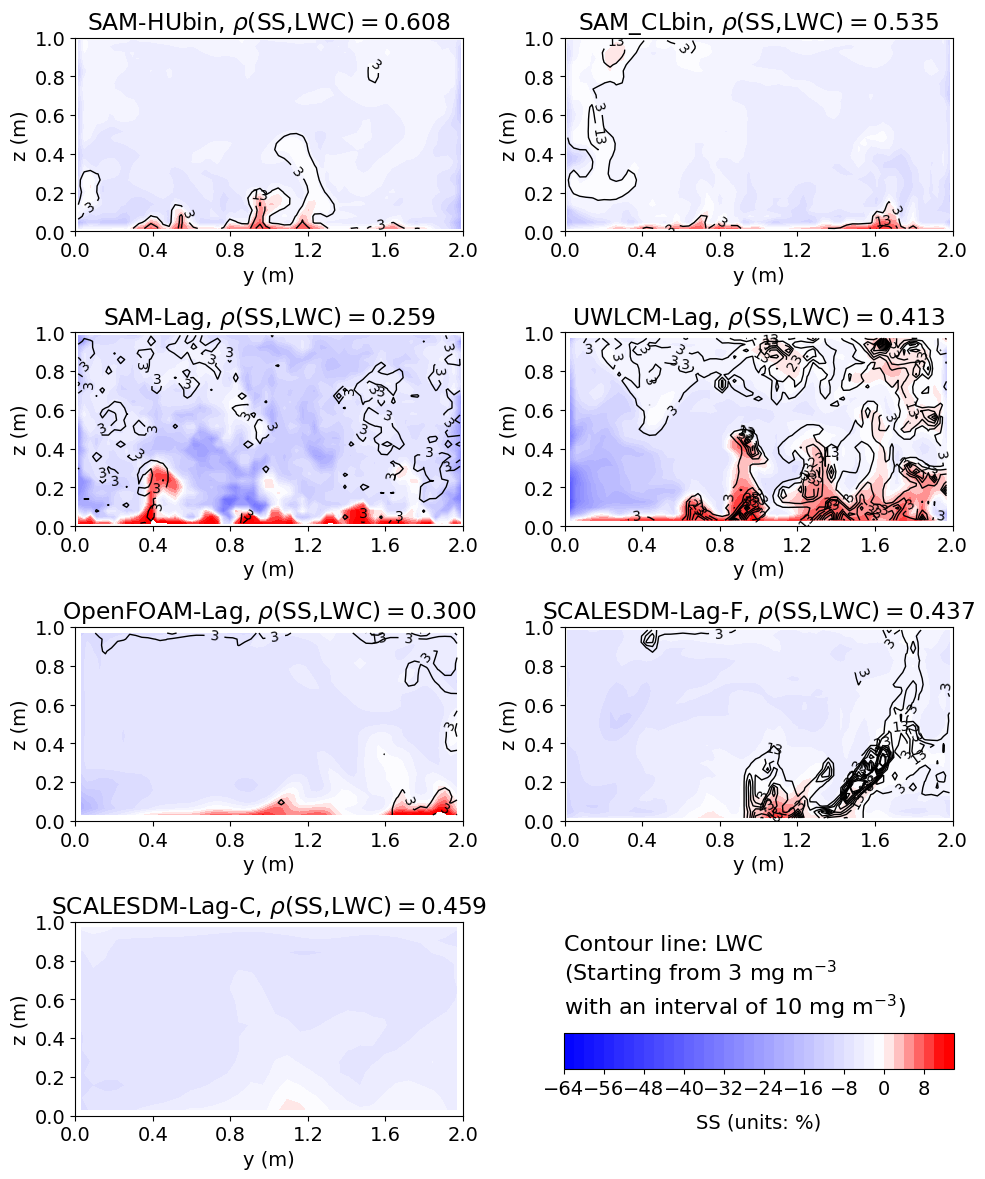}
    \caption{The slices of supersaturation with respect to liquid (shading) and liquid water content (contour lines) as the ice injection rate is 10 cm$^{-3}$ min$^{-1}$. The locations of the slices are chosen to include the maximum supersaturation. The correlation coefficient ($\rho$) between supersaturation and liquid water content in the entire domain is shown in the title of each subpanel.}
    \label{fig:SS_slices}
\end{figure*}

Fig.~\ref{fig:SS_slices} illustrates vertical slices from the LES models. Supersaturation is highest near the bottom and gradually mixed upward into the bulk area by turbulence. Although SCALE-SDM-Lag-C is able to produce a localized supersaturation region near the center of the bottom, the area is very small and insufficient to maintain the droplets. Additionally, the correlation coefficients between supersaturation and liquid water content in the models employing bin microphysics are all higher than those using the Lagrangian approach, implying that Lagrangian particles are less coupled with the supersaturation fields.

\section{Conclusions}

This study presents a comprehensive intercomparison of ten mixed-phase cloud simulations of the Pi Cloud Chamber. The primary objectives are to evaluate each model's ability to reproduce the observed quasi-steady mixed-phase cloud state, understand the convergence or discrepancy among models, and identify the factors influencing cloud glaciation processes under controlled laboratory conditions. 
Key findings from the intercomparison include:
\begin{itemize} 
\item All models successfully reproduce the observed mean droplet radius in the liquid-phase stage (i.e., before ice injection) by tuning model parameters, such as (for simulating the entire domain) side-wall wetness and CCN injection rates, or (for simulating only the core region without the walls) supersaturation forcing and particle settling rates. 
\item The DNS and statistical models, which simulate only the core region of the chamber, require external forcing to account for wall effects. Despite simplified assumptions, such as homogeneous isotropic turbulence in the core region and directly solving for the supersaturation equation instead of both temperature and water vapor, both models effectively captured the key glaciation trends when appropriately forced, demonstrating the utility of reduced-complexity approaches. 
\item Greater inter-model variability is observed in predicting liquid-phase processes than in ice microphysics, which leads to model divergence in glaciation process. 
\item Supersaturation plays a central role in cloud evolution. All models show decreasing mean supersaturation with increasing ice number concentration; however, the timing and magnitude of this decrease varies across models. In most cases, even under negative mean supersaturation, droplets persist due to strong supersaturation fluctuations, especially near the bottom of the chamber where moisture flux is strongest. 
\item Most LES models do not achieve full glaciation within the tested ice injection range. The bulk model and coarse-resolution LES reach full glaciation more readily due to enhanced mixing and reduced spatial heterogeneity. The statistical and DNS models reach full glaciation because they excluded the near-wall region. 
\item Bin microphysics shows stronger coupling between supersaturation and liquid water content, while Lagrangian schemes produce larger ice particles and lower number concentrations, likely due to weaker coupling and greater exposure to supersaturation variability. 
\end{itemize}
These findings highlight the importance of resolving fine-scale turbulence and supersaturation variability for accurately simulating mixed-phase cloud processes. They also demonstrate that differences in model configurations, such as grid resolution, domain extent, and microphysical schemes, can significantly influences the microphysical states. Furthermore, this cloud chamber-based modeling study underscore the value of laboratory facilities in providing reproducible, well-constrained environments for model evaluation. Additionally, the models not only help to interpret laboratory results (e.g., Lagrangianly tracking particles of different phases, explaining why achieving full glaciation is so challenging) but also assist in identifying key locations for sensor placement to enhance laboratory experiments. Lastly, this intercomparison study lays the groundwork for future efforts to improve the representation of mixed-phase clouds in atmospheric models, ultimately contributing to more accurate weather and climate predictions.

\section*{Conflict of Interest declaration}
The authors declare there are no conflicts of interest for this manuscript.

\section*{Data and Code Availability}
The post-processed data are available at Zenodo via the following link: \url{https://doi.org/10.5281/zenodo.17428022} \cite{Wang2025Zenodo}. The models used for this work are available as follows:
\begin{enumerate}

\item The bulk model used in this study is openly available on Zenodo at \url{https://zenodo.org/records/17858499} \cite{Krueger_2025_17858499}. The Zenodo record includes the full source code, documentation, and an example simulation.

\item The code for the DNS model is available at \url{https://zenodo.org/records/17865055} \cite{Sardina_2025_17865055}.

\item The code for the Statistical model is available at \url{https://zenodo.org/records/17865470} \cite{Sarnitsky_2025_17865470}.

\item The SAM model was developed by Dr. Marat Khairoutdinov from Stony Brook University \cite{Khairoutdinov2003JAS} and is publicly available at \url{http://rossby.msrc.sunysb.edu/SAM.html}. 

\item OpenFOAM is an open source available at \url{https://www.openfoam.com/}. The version of OpenFOAM used for this work is OpenFOAM-dev 20200614 \cite{OpenFOAM}. 

\item The UWLCM model and its dependencies, libcloudph++ and libmpdata++, are open-source available to download at Zenodo. The versions used in this studies are: 
(1) UWLCM 1.1p \cite{uwlcm}, 
(2) libcloudph++ 2.1.1p: \cite{UWLCM_libcloudph}, 
(3) libmpdata++ 1.2.1p: \cite{UWLCM_libmpdata}.

\item SCALE-SDM: Version SCALE 5.2.6-SDM 2.2.2 \cite{SCALE,SDM} with modifications for sidewalls and particle injection is available at \url{https://doi.org/10.5281/zenodo.17791338} \cite{SCALE-SDM}
\end{enumerate}

\acknowledgments

We thank Neel Desai from San Jose State University and Susanne Glienke from Pacific Northwest National Laboratory (PNNL) for the information and discussion related to the hologram used for the observation. 

This research is supported by the U.S. Department of Energy (DOE) Office of Science (SC) Atmospheric System Research (ASR) project at PNNL. PNNL is operated for the DOE by Battelle Memorial Institute under Contract DE-AC05-76 RL01830. This research used resources of the National Energy Research Scientific Computing Center (NERSC), a DOE SC User Facility located at Lawrence Berkeley National Laboratory, operated under Contract No.~DE-AC02-05CH11231 using NERSC awards BER-ERCAP0033002.

Chen’s work is based upon work supported by the NSF National Center for Atmospheric Research, which is a major facility sponsored by the U.S. National Science Foundation under Cooperative Agreement No.~1852977. 

Krueger's research was supported by National Science Foundation Grant AGS-2133229.

A. Makulska and P. Dziekan gratefully acknowledge Polish high-performance computing infrastructure PLGrid (HPC Centers: ACK Cyfronet AGH) for providing computer facilities and support within computational grant No.~PLG/2023/016820.

F.~Hoffmann appreciates support from the Emmy Noether program of the German Research Foundation (DFG) under grant HO 6588/1-1. 

F.~Yang was funded by DOE as part of the Atmospheric System Research (ASR) program under contract DE-SC0012704. 

To produce SCALE-SDM results, K.~Enokido and S.~Shima used the computational resources of the supercomputer Grand Chariot provided by Hokkaido University through the HPCI System Research Project (project IDs: hp200078, hp210059, hp220062, hp230166, and hp240151) and the computer facilities of the Center for Cooperative Work on Data science and Computational science, University of Hyogo. K.~Enokido and S.~Shima were supported by JSPS KAKENHI, Grant 20H00225 and 23H00149; and JST (Moonshot R and D) (Grant JPMJMS2286 and JPMJMS2283).

B.~Mehlig and G.~Sarnitsky were supported by Vetenskapsr\aa{}det (grant No.~2021-4452). G.~Sardina was supported by Vetenskapsr\aa{}det (grant No.~2023-2026) and  by ERC grant MixClouds 101126050 funded by the European Union. The computations were enabled by resources provided by the National Academic Infrastructure for Supercomputing in Sweden (NAISS), partially funded by the Swedish Research Council through grant agreement No.~2022-06725.
Views and opinions expressed are, however, those of the author(s) only and do not necessarily reflect those of the European Union or the European Research Council Executive Agency. Neither the European Union nor the granting authority can be held responsible for them.

R.~Shaw was supported by US National Science Foundation grant AGS-2113060.

Lastly, following the Committee on Publication Ethics (COPE) recommendations  for transparency, we acknowledge the use of large-language model which aided in generating the first draft of data processing code, proofreading for spelling and grammar, and providing suggestions for contextual improvement.

%
%

\bibliography{Ref.bib}

@STRING{QJRMS     = "Quart.\ J.\ Roy.\ Meteor.\ Soc."}

@STRING{MA        = "Meteor.\ Appl."}

@STRING{AN        = "Astrophys.\ Norv."}

@STRING{AIP 	= "Amer. Inst. Phys."}

@Article{Nishizawa2015,
AUTHOR = {Nishizawa, S. and Yashiro, H. and Sato, Y. and Miyamoto, Y. and Tomita, H.},
TITLE = {Influence of grid aspect ratio on planetary boundary layer turbulence in large-eddy simulations},
JOURNAL = {Geoscientific Model Development},
VOLUME = {8},
YEAR = {2015},
NUMBER = {10},
PAGES = {3393--3419},
DOI = {10.5194/gmd-8-3393-2015}
}

@article{Sato2015,
author = {Sato, Yousuke and Nishizawa, Seiya and Yashiro, Hisashi and Miyamoto, Yoshiaki and Kajikawa, Yoshiyuki and Tomita, Hirofumi},
doi = {10.1186/s40645-015-0053-6},
issn = {21974284},
journal = {Progress in Earth and Planetary Science},
month = {dec},
number = {1},
publisher = {Springer Berlin Heidelberg},
title = {{Impacts of cloud microphysics on trade wind cumulus: which cloud microphysics processes contribute to the diversity in a large eddy simulation?}},
volume = {2},
year = {2015}
}

@inbook{Koren1993,
title = "A robust upwind discretization method for advection, diffusion and source terms",
author = "B. Koren",
year = "1993",
language = "English",
series = "Notes on Numerical Fluid Mechanics",
publisher = "Vieweg",
pages = "117--138",
editor = "C.B. Vreugdenhil and B. Koren",
booktitle = "Numerical Methods for Advection-Diffusion Problems",
address = "Germany",
}

@article {Wicker2002,
      author = "Louis J. Wicker and William C. Skamarock",
      title = "Time-Splitting Methods for Elastic Models Using Forward Time Schemes",
      journal = "Monthly Weather Review",
      year = "2002",
      publisher = "American Meteorological Society",
      address = "Boston MA, USA",
      volume = "130",
      number = "8",
      doi = "10.1175/1520-0493(2002)130<2088:TSMFEM>2.0.CO;2",
      pages=      "2088 - 2097",
}

@article{Scotti1993,
author = {Scotti, Alberto and Meneveau, Charles and Lilly, Douglas K.},
doi = {10.1063/1.858537},
issn = {0899-8213},
journal = {Physics of Fluids A: Fluid Dynamics},
month = {sep},
number = {9},
pages = {2306--2308},
title = {{Generalized Smagorinsky model for anisotropic grids}},
volume = {5},
year = {1993}
}

@article{Smagorinsky1963,
author = {Smagorinsky, J},
doi = {10.1175/1520-0493(1963)091<0099:GCEWTP>2.3.CO;2},
issn = {0027-0644},
journal = {Monthly Weather Review},
month = {mar},
number = {3},
pages = {99--164},
publisher = {American Meteorological Society},
title = {{GENERAL CIRCULATION EXPERIMENTS WITH THE PRIMITIVE EQUATIONS}},
volume = {91},
year = {1963}
}

@article{Brown1994,
author = {Brown, A. R. and Derbyshire, S. H. and Mason, P. J.},
title = {Large-eddy simulation of stable atmospheric boundary layers with a revised stochastic subgrid model},
journal = {Quarterly Journal of the Royal Meteorological Society},
volume = {120},
number = {520},
pages = {1485-1512},
doi = {https://doi.org/10.1002/qj.49712052004},
eprint = {https://rmets.onlinelibrary.wiley.com/doi/pdf/10.1002/qj.49712052004},
year = {1994}
}

@misc{OpenFOAM,
title={OpenFOAM-dev 20200614},
author={{OpenCFD Ltd}},
version={20200614},
year={2020},
url={https://dl.openfoam.org/ubuntu/551
dists/xenial/dev/binary-amd64/openfoam-dev 20200614 amd64.deb},
note={software}
}

@misc{UWLCM_libcloudph,
title={libcloudph++ - a cloud (micro)physics library, version 2.1.1p},
author={Dziekan, Piotr  and Arabas, Sylwester  and Jaruga, Anna  and Waruszewski, Maciej  and Jarecka, Dorota  and Makulska, Agnieszka and Zmijewski, Piotr and Singer, Clare  and Badger, Codacy},
year={2025},
doi={https://doi.org/10.5281/zenodo.14965763},
note={software}
}

@misc{UWLCM_libmpdata,
title={libmpdata++ - a library of parallel MPDATA-based solvers for systems of generalised transport equations, version 1.2.1p},
author={Arabas, Sylwester  and Dziekan, Piotr  and Waruszewski, Maciej  and Jaruga, Anna  and Jarecka, Dorota  and Badger, Codacy  and Singer, Clare },
year={2025},
doi={https://doi.org/10.5281/zenodo.14965764},
note={software}
}

@misc{SCALE,
title={SCALE 5.2.6},
url={https://scale.riken.jp/archives/scale-5.2.6.tar.gz},
author={Nishizawa, S. and Yashiro, H. and Yamaura, T. and Adachi, Sachiho and A., Yoshida, R. and Sato, Y. and Sueki, K. and Matsushima, T. and Kawai, Y. and Yanase, T. and Tomita, H},
year={2018},
note={software}
}

@misc{SDM,
title={Super-droplet Method (SDM) 2.2.2},
author={Shima, Shin-ichiro},
doi={https://doi.org/10.5281/zenodo.3483650},
year={2020},
note={software}
}

@misc{SCALE-SDM,
title={SCALE-SDM 5.2.6-2.2.2 for ICMW 2024 Pi Chamber Mixed-Phase Cloud Simulation Case},
author={Shima, Shin-ichiro and Enokido, Kotaro},
doi={https://doi.org/10.5281/zenodo.17791338},
year={2025},
note={software}
}

@Article{Shima2020,
AUTHOR = {Shima, S. and Sato, Y. and Hashimoto, A. and Misumi, R.},
TITLE = {Predicting the morphology of ice particles in deep convection using the super-droplet method:
development and evaluation of SCALE-SDM 0.2.5-2.2.0, -2.2.1, and -2.2.2},
JOURNAL = {Geoscientific Model Development},
VOLUME = {13},
YEAR = {2020},
NUMBER = {9},
PAGES = {4107--4157},
DOI = {10.5194/gmd-13-4107-2020}
}

@article{grabowski2018,
	author = {Grabowski, Wojciech W and Dziekan, Piotr and Pawlowska, Hanna},
	journal = {Geosci. Model Dev.},
	number = {1},
	pages = {103},
	publisher = {Copernicus GmbH},
	title = {Lagrangian condensation microphysics with Twomey CCN activation},
	volume = {11},
	year = {2018}}

@article{hoffmann2015,
	author = {Hoffmann, F and Raasch, S. and Noh, Y.},
	journal = {Atmos. Res.},
	pages = {43--57},
	title = {Entrainment of aerosols and their activation in a shallow cumulus cloud studied with a coupled {LCM}-{LES} approach},
	volume = {156},
	year = {2015}}

@article{hoffmann2019,
	author = {Hoffmann, F and Yamaguchi, Takanobu and Feingold, Graham},
	doi = {10.1175/JAS-D-18-0087.1},
	journal = {J. Atmos. Sci.},
	number = {1},
	pages = {113--133},
	title = {{Inhomogeneous mixing in Lagrangian cloud models: Effects on the production of precipitation embryos}},
	volume = {76},
	year = {2019}}

@article{hoffmann2020,
	author = {Hoffmann, Fabian},
	journal = {J. Atmos. Sci.},
	number = {6},
	pages = {2279--2296},
	title = {Effects of Entrainment and Mixing on the {Wegener--Bergeron--Findeisen} Process},
	volume = {77},
	year = {2020}}

@article{Desai2018JAS,
  title={Influence of microphysical variability on stochastic condensation in a turbulent laboratory cloud},
  author={Desai, N and Chandrakar, Kamal Kant and Chang, K and Cantrell, W and Shaw, RA},
  journal={Journal of the Atmospheric Sciences},
  volume={75},
  number={1},
  pages={189--201},
  year={2018}
}

@incollection{Arakawa1977,
  title     = {Computational Design of the Basic Dynamical Processes of the UCLA General Circulation Model},
  series    = {Methods in Computational Physics: Advances in Research and Applications},
  publisher = {Elsevier},
  volume    = {17},
  pages     = {173 - 265},
  year      = {1977},
  booktitle = {General Circulation Models of the Atmosphere},
  issn      = {0076-6860},
  doi       = {https://doi.org/10.1016/B978-0-12-460817-7.50009-4},
  author    = {Akio Arakawa and Vivian R. Lamb}
}

@book{Kays1980,
  title={Convective heat and mass transfer},
  author={Kays, William Morrow and Crawford, Michael E and Weigand, Bernhard},
  volume={4},
  year={1980},
  publisher={McGraw-Hill New York}
}

@article{MacMillan2022PRF,
  title={Direct numerical simulation of turbulence and microphysics in the Pi Chamber},
  author={MacMillan, Theodore and Shaw, Raymond A and Cantrell, Will H and Richter, David H},
  journal={Physical Review Fluids},
  volume={7},
  number={2},
  pages={020501},
  year={2022},
  publisher={APS}
}

@article{Yang2022JAMES,
  title={Large-eddy simulations of a convection cloud chamber: Sensitivity to bin microphysics and advection},
  author={Yang, Fan and Ovchinnikov, Mikhail and Thomas, Subin and Khain, Alexander and McGraw, Robert and Shaw, Raymond A and Vogelmann, Andrew M},
  journal={Journal of Advances in Modeling Earth Systems},
  volume={14},
  number={5},
  pages={e2021MS002895},
  year={2022},
  publisher={Wiley Online Library}
}

@article{Yang2023JAMES,
  title={An Intercomparison of Large-Eddy Simulations of a Convection Cloud Chamber Using Haze-Capable Bin and Lagrangian Cloud Microphysics Schemes},
  author={Yang, Fan and Hoffmann, Fabian and Shaw, Raymond A and Ovchinnikov, Mikhail and Vogelmann, Andrew M},
  journal={Journal of Advances in Modeling Earth Systems},
  volume={15},
  number={5},
  pages={e2022MS003270},
  year={2023},
  publisher={Wiley Online Library}
}

@software{Krueger_2025_17858499,
  author       = {Krueger, Steven},
  title        = {Bulk Model Code for 'A Model Intercomparison Study
                   of Mixed-Phase Clouds in a Laboratory Chamber'
                  },
  month        = dec,
  year         = 2025,
  publisher    = {Zenodo},
  doi          = {10.5281/zenodo.17858499},
  url          = {https://doi.org/10.5281/zenodo.17858499},
  swhid        = {swh:1:dir:66dfb405ba5d857d62b8a16909a8c778cc726e4e
                   ;origin=https://doi.org/10.5281/zenodo.17858498;vi
                   sit=swh:1:snp:94fc898ee2a56906606495c437dba2799d13
                   4455;anchor=swh:1:rel:f29f02435075446a6030860747ed
                   f67f6815abe9;path=/
                  },
}

@software{Sardina_2025_17865055,
  author       = {Sardina, Gaetano},
  title        = {DNS-HI Homogeneous Isotropic flow},
  month        = dec,
  year         = 2025,
  publisher    = {Zenodo},
  doi          = {10.5281/zenodo.17865055},
  url          = {https://doi.org/10.5281/zenodo.17865055},
}

@misc{Sarnitsky_2025_17865470,
  author       = {Sarnitsky, Grigory},
  title        = {Statistical Model for the Pi Chamber Core Region},
  month        = dec,
  year         = 2025,
  publisher    = {Zenodo},
  doi          = {10.5281/zenodo.17865470},
  url          = {https://doi.org/10.5281/zenodo.17865470},
}

@article{Chandrakar2016PNAS,
  author = {Kamal Kant Chandrakar  and Will Cantrell  and Kelken Chang  and David Ciochetto  and Dennis Niedermeier  and Mikhail Ovchinnikov  and Raymond A. Shaw  and Fan Yang },
  title = {Aerosol indirect effect from turbulence-induced broadening of cloud-droplet size distributions},
  journal = {Proceedings of the National Academy of Sciences},
  volume = {113},
  number = {50},
  pages = {14243-14248},
  year = {2016},
  doi = {10.1073/pnas.1612686113}
}

@article {Chang2016BAMS,
  author = "K. Chang and J. Bench and M. Brege and W. Cantrell and K. Chandrakar and D. Ciochetto and C. Mazzoleni and L. R. Mazzoleni and D. Niedermeier and R. A. Shaw",
  title = "A Laboratory Facility to Study Gas-Aerosol-Cloud Interactions in a Turbulent Environment: The $\Pi$ Chamber",
  journal = "Bulletin of the American Meteorological Society",
  year = "2016",
  publisher = "American Meteorological Society",
  address = "Boston MA, USA",
  volume = "97",
  number = "12",
  doi = "10.1175/BAMS-D-15-00203.1",
  pages=      "2343 - 2358"
}

@article{Deardorff1980BLM,
  title     = {Stratocumulus-capped mixed layers derived from a three-dimensional model},
  author    = {Deardorff, J. W.},
  journal   = {Boundary-Layer Meteorol.},
  volume    = {18},
  pages     = {495--527},
  year      = {1980},
  publisher = {Springer}
}

@article{Fan2011,
   author = {Fan, J. W. and Ghan, S. and Ovchinnikov, M. and Liu, X. H. and Rasch, P. J. and Korolev, A.},
   title = {{Representation of Arctic mixed-phase clouds and the Wegener-Bergeron-Findeisen process in climate models: Perspectives from a cloud-resolving study}},
   journal = {Journal of Geophysical Research-Atmospheres},
   volume = {116},
   pages = {17},
   DOI = {10.1029/2010jd015375},
   year = {2011}
}

@article {Khain2004JAS,
  author = "A. Khain and A. Pokrovsky and M. Pinsky and A. Seifert and V. Phillips",
  title = "Simulation of Effects of Atmospheric Aerosols on Deep Turbulent Convective Clouds Using a Spectral Microphysics Mixed-Phase Cumulus Cloud Model. Part I: Model Description and Possible Applications",
  journal = "Journal of the Atmospheric Sciences",
  year = "2004",
  publisher = "American Meteorological Society",
  address = "Boston MA, USA",
  volume = "61",
  number = "24",
  doi = "10.1175/JAS-3350.1",
  pages=      "2963 - 2982"
}

@article {Khairoutdinov2003JAS,
  author = "Marat F. Khairoutdinov and David A. Randall",
  title = "Cloud Resolving Modeling of the ARM Summer 1997 IOP: Model Formulation, Results, Uncertainties, and Sensitivities",
  journal = "Journal of the Atmospheric Sciences",
  year = "2003",
  publisher = "American Meteorological Society",
  address = "Boston MA, USA",
  volume = "60",
  number = "4",
  doi = "10.1175/1520-0469(2003)060<0607:CRMOTA>2.0.CO;2",
  pages=      "607 - 625"
}

@article{Korolev2017MM,
	title = {Mixed-{Phase} {Clouds}: {Progress} and {Challenges}},
	volume = {58},
	copyright = {http://www.ametsoc.org/PUBSReuseLicenses},
	issn = {0065-9401},
	shorttitle = {Mixed-{Phase} {Clouds}},
	doi = {10.1175/amsmonographs-d-17-0001.1},
	language = {en},
	urldate = {2025-07-14},
	journal = {Meteorological Monographs},
	author = {Korolev, A. and McFarquhar, G. and Field, P. R. and Franklin, C. and Lawson, P. and Wang, Z. and Williams, E. and Abel, S. J. and Axisa, D. and Borrmann, S. and Crosier, J. and Fugal, J. and Krämer, M. and Lohmann, U. and Schlenczek, O. and Schnaiter, M. and Wendisch, M.},
	month = jan,
	year = {2017},
	note = {Publisher: American Meteorological Society},
	pages = {5.1--5.50},
}

@misc{Wang2025Zenodo,
	title = {Model {Output} from the {Mixed}-{Phase} {Cloud} {Chamber} {Intercomparison} at {ICMW} 2024},
	copyright = {Creative Commons Attribution 4.0 International},
	doi = {10.5281/ZENODO.15811058},
	urldate = {2025-07-14},
	publisher = {Zenodo},
	author = {Wang, Aaron and Chen, Sisi and Krueger, Steve and Dziekan, Piotr and Enokido, Kotaro and Hoffmann, Fabian and Makulska, Agnieszka and Mehlig, Bernhard and Sardina, Gaetano and Sarnitsky, Grigory and Schmalfuß, Silvio and Shima, Shin-ichiro and Yang, Fan and Ovchinnikov, Mikhail and Shaw, Raymond},
	month = jul,
	year = {2025},
}

@article{Smolarkiewicz1990JCP,
  title = {The multidimensional positive definite advection transport algorithm: nonoscillatory option},
  journal = {Journal of Computational Physics},
  volume = {86},
  number = {2},
  pages = {355-375},
  year = {1990},
  issn = {0021-9991},
  doi = {10.1016/0021-9991(90)90105-A},
  author = {Piotr K Smolarkiewicz and Wojciech W Grabowski}
}

@article{Thomas2019JAMES,
  author = {Thomas, Subin and Ovchinnikov, Mikhail and Yang, Fan and van der Voort, Dennis and Cantrell, Will and Krueger, Steven K. and Shaw, Raymond A.},
  title = {Scaling of an Atmospheric Model to Simulate Turbulence and Cloud Microphysics in the Pi Chamber},
  journal = {Journal of Advances in Modeling Earth Systems},
  volume = {11},
  number = {7},
  pages = {1981-1994},
  doi = {10.1029/2019MS001670},
  year = {2019}
}

@article{wang2024entrainment,
  title={The Dual Nature of Entrainment-Mixing Signatures Revealed Through Large-Eddy Simulations of a Convection-Cloud Chamber},
  author={Wang, Aaron and Ovchinnikov, Mikhail and Yang, Fan and Cantrell, Will and Yeom, Jaemin and Shaw, Raymond A},
  journal={Journal of the Atmospheric Sciences},
  volume={81},
  number={12},
  pages={2017--2039},
  year={2024},
  publisher={American Meteorological Society}
}

@article{Yang2025ACP,
  title={Microphysics regimes due to haze--cloud interactions: cloud oscillation and cloud collapse},
  author={Yang, Fan and Sadi, Hamed Fahandezh and Shaw, Raymond A and Hoffmann, Fabian and Hou, Pei and Wang, Aaron and Ovchinnikov, Mikhail},
  journal={Atmospheric Chemistry and Physics},
  volume={25},
  number={6},
  pages={3785--3806},
  year={2025},
  publisher={Copernicus Publications G{\"o}ttingen, Germany}
}

@article{Wang2025PoF,
  title={An intercomparison of wall fluxes in a turbulent thermal convection chamber: Direct numerical simulations and wall-modeled large-eddy simulations enhanced by machine learning},
  author={Wang, Aaron and Schmalfu{\ss}, Silvio and Chandrakar, Kamal Kant and Kia, Hadi Zanganeh and Yang, Fan and Ovchinnikov, Mikhail and Shaw, Raymond A and Choi, Yunsoo},
  journal={Physics of Fluids},
  volume={37},
  number={4},
  year={2025},
  publisher={AIP Publishing}
}

@Article{Wang2024ACP,
    author = {Wang, A. and Krueger, S. and Chen, S. and Ovchinnikov, M. and Cantrell, W. and Shaw, R. A.},
    doi = {10.5194/acp-24-10245-2024},
    journal = {Atmospheric Chemistry and Physics},
    number = {18},
    pages = {10245--10260},
    title = {Glaciation of mixed-phase clouds: insights from bulk model and bin-microphysics large-eddy simulation informed by laboratory experiment},
    volume = {24},
    year = {2024}
}

@article{xue2022,
  title={Progress and challenges in modeling dynamics--microphysics interactions: From the pi chamber to monsoon convection},
  author={Xue, Lulin and Bera, Sudarsan and Chen, Sisi and Choudhary, Harish and Dixit, Shivsai and Grabowski, Wojciech W and Jayakumar, Sandeep and Krueger, Steven and Kulkarni, Gayatri and Lasher-Trapp, Sonia and others},
  journal={Bulletin of the American Meteorological Society},
  volume={103},
  number={5},
  pages={E1413--E1420},
  year={2022}
}

@article{Desai2019GRL,
  title={Aerosol-mediated glaciation of mixed-phase clouds: Steady-state laboratory measurements},
  author={Desai, Neel and Chandrakar, KK and Kinney, G and Cantrell, W and Shaw, RA},
  journal={Geophysical Research Letters},
  volume={46},
  number={15},
  pages={9154--9162},
  year={2019},
  publisher={Wiley Online Library}
}

@article{Wang2024JAS,
  title={An Investigation of {LES} Wall Modeling for {Rayleigh--B{\'e}nard} Convection via Interpretable and Physics-Aware Feedforward Neural Networks with {DNS}},
  author={Wang, Aaron and Yang, Xiang IA and Ovchinnikov, Mikhail},
  journal={Journal of the Atmospheric Sciences},
  volume={81},
  number={2},
  pages={435--458},
  year={2024},
  publisher={American Meteorological Society}
}

@article{Shima2009QJRMS,
  title={The super-droplet method for the numerical simulation of clouds and precipitation: A particle-based and probabilistic microphysics model coupled with a non-hydrostatic model},
  author={Shima, Shin-ichiro and Kusano, Kanya and Kawano, Akio and Sugiyama, Tooru and Kawahara, Shintaro},
  journal={Quarterly Journal of the Royal Meteorological Society},
  volume={135},
  number={642},
  pages={1307--1320},
  year={2009},
  doi = {https://doi.org/10.1002/qj.441},
  eprint = {https://rmets.onlinelibrary.wiley.com/doi/pdf/10.1002/qj.441},
  publisher={Wiley Online Library}
}

@article{Wang2024JAMES,
  title={Designing a Convection-Cloud Chamber for Collision-Coalescence Using Large-Eddy Simulation With Bin Microphysics},
  author={Wang, Aaron and Ovchinnikov, Mikhail and Yang, Fan and Schmalfuss, Silvio and Shaw, Raymond A},
  journal={Journal of Advances in Modeling Earth Systems},
  volume={16},
  number={1},
  pages={e2023MS003734},
  year={2024},
  publisher={Wiley Online Library}
}

@article{Arakawa1977GCMA,
  title={Computational design of the basic dynamical processes of the UCLA general circulation model},
  author={Arakawa, Akio and Lamb, Vivian R},
  journal={General circulation models of the atmosphere},
  volume={17},
  number={Supplement C},
  pages={173--265},
  year={1977},
  publisher={Methods in computational Physics}
}

@Article{Krueger2020ACP,
AUTHOR = {Krueger, S. K.},
TITLE = {Technical note: Equilibrium droplet size distributions in a turbulent cloud chamber with uniform supersaturation},
JOURNAL = {Atmospheric Chemistry and Physics},
VOLUME = {20},
YEAR = {2020},
NUMBER = {13},
PAGES = {7895--7909},
DOI = {10.5194/acp-20-7895-2020}
}

@article{Morrison2011,
   author = {Morrison, H. and Zuidema, P. and Ackerman, A. S. and Avramov, A. and de Boer, G. and Fan, J. W. and Fridlind, A. M. and Hashino, T. and Harrington, J. Y. and Luo, Y. L. and Ovchinnikov, M. and Shipway, B.},
   title = {{Intercomparison of cloud model simulations of Arctic mixed-phase boundary layer clouds observed during SHEBA/FIRE-ACE}},
   journal = {Journal of Advances in Modeling Earth Systems},
   volume = {3},
   DOI = {10.1029/2011ms000066},
   year = {2011}
}

@article{chen1994simulation,
  title={Simulation of cloud microphysical and chemical processes using a multicomponent framework. Part I: Description of the microphysical model},
  author={Chen, Jen-Ping and Lamb, Dennis},
  journal={Journal of Atmospheric Sciences},
  volume={51},
  number={18},
  pages={2613--2630},
  year={1994},
  doi={10.1175/1520-0469(1994)051<2613:SOCMAC>2.0.CO;2}
}

@article{sarnitsky2025does,
  title = {Does small-scale turbulence matter for ice growth in mixed-phase clouds?},
  author = {Sarnitsky, Grigory and Sardina, Gaetano and Svensson, Gunilla and Pumir, Alain and Hoffmann, Fabian and Mehlig, Bernhard},
  journal = {Physical Review Fluids},
  volume = {10},
  number = {5},
  pages = {053803},
  year = {2025},
  publisher = {APS}
}

@article{fries2021key,
  title={Key parameters for droplet evaporation and mixing at the cloud edge},
  author={Fries, J and Sardina, G and Svensson, G and Mehlig, B},
  journal={QJRMS},
  year={2021},
  pages = {2160--2172},
  volume = {147},
  publisher={Wiley Online Library}
}

@article{sardina2015continuous,
  title={Continuous growth of droplet size variance due to condensation in turbulent clouds},
  author={Sardina, G and Picano, F and Brandt, L and Caballero, R},
  journal={Phys. Rev. Lett.},
  volume={115},
  number={18},
  pages={184501},
  year={2015},
  publisher={APS}
}

@article{pope1991mapping,
  title={Mapping closures for turbulent mixing and reaction},
  author={Pope, SB},
  journal={Theor. Comput. Fluid Dyn.},
  volume={2},
  number={5},
  pages={255--270},
  year={1991},
  publisher={Springer}
}

@article{saito19,
  title={Broadening of cloud droplet size distributions by condensation in turbulence},
  author={Saito, I and Gotoh, T and Watanabe, T},
  journal={Journal of the Meteorological Society of Japan. Ser. II},
  volume={97},
  number={4},
  pages={867--891},
  year={2019},
  publisher={Meteorological Society of Japan}
}

@article{gosman1983aspects,
  title={Aspects of computer simulation of liquid-fueled combustors},
  author={Gosman, AD and Loannides, E},
  journal={Journal of energy},
  volume={7},
  number={6},
  pages={482--490},
  year={1983}
}

@article{chen2023mixed,
  title={Mixed-phase direct numerical simulation: ice growth in cloud-top generating cells},
  author={Chen, Sisi and Xue, Lulin and Tessendorf, Sarah and Ikeda, Kyoko and Weeks, Courtney and Rasmussen, Roy and Kunkel, Melvin and Blestrud, Derek and Parkinson, Shaun and Meadows, Melinda and others},
  journal={Atmospheric Chemistry and Physics},
  volume={23},
  number={9},
  pages={5217--5231},
  year={2023},
  publisher={Copernicus Publications G{\"o}ttingen, Germany}
}

@article{Dziekan2019,
author = {Dziekan, P. and Waruszewski, M. and Pawlowska, H.},
title = {University of Warsaw Lagrangian Cloud Model ({UWLCM}) 1.0: a modern large-eddy simulation tool for warm cloud modeling with Lagrangian microphysics},
journal = {Geoscientific Model Development},
volume = {12},
year = {2019},
number = {6},
pages = {2587--2606},
doi = {10.5194/gmd-12-2587-2019}
}

@article{Arabas2015,
author = {Arabas, S. and Jaruga, A. and Pawlowska, H. and Grabowski, W. W.},
title = {libcloudph++ 1.0: a single-moment bulk, double-moment bulk, and particle-based warm-rain microphysics library in C++},
journal = {Geoscientific Model Development},
volume = {8},
year = {2015},
number = {6},
pages = {1677--1707},
doi = {10.5194/gmd-8-1677-2015}
}

@article{schiller1933uber,
  title={Uber die grundlegenden Berechnungen bei der Schwerkraftaufbereitung},
  author={Schiller, L. and Naumann, A.},
  journal={Z. VDI},
  volume={77},
  pages={318--321},
  year={1933}
}

@article{ranz1952evaporation,
  title={The evaporation from drops},
  author={Ranz, WE and Marshall, WR},
  journal={Chem. Eng. Prog},
  volume={48},
  number={3},
  pages={141--146},
  year={1952}
}

@article{cunningham1910velocity,
  title={On the velocity of steady fall of spherical particles through fluid medium},
  author={Cunningham, Ebenezer},
  journal={Proceedings of the Royal Society of London. Series A, Containing Papers of a Mathematical and Physical Character},
  volume={83},
  number={563},
  pages={357--365},
  year={1910},
  publisher={The Royal Society London}
}

@software{UWLCM,
  author       = {Piotr Dziekan and
                  Clare Singer and
                  Maciej Waruszewski and
                  Anna Jaruga and
                  Piotr Żmijewski},
  title        = {{University of Warsaw Lagrangian Cloud Model source code}},
  year         = 2025,
  publisher    = {Zenodo},
  version      = {v2.1p},
  doi          = {10.5281/zenodo.15623184},
}

@article{Chen2025,
  author       = {Sisi Chen and Steven K. Krueger and Piotr Dziekan and Kotaro Enokido and Theodore Macmillan and David Richter and Silvio Schmalfuß and Shin-ichiro Shima and Fan Yang and Jesse C. Anderson and Will Cantrell and Dennis Niedermeier and Raymond A. Shaw and Frank Stratmann},
  title        = {A Model Inter-comparison Study of Aerosol‐Cloud‐Turbulence Interactions in a Cloud Chamber: 1. Model Results},
  journal      = {Journal of Advances in Modeling Earth Systems},
  year         = {2025},
  doi          = {10.1029/2024MS004562},
}

@article{Villanueva2022,
doi = {10.1088/1748-9326/aca16d},
year = {2022},
month = {nov},
publisher = {IOP Publishing},
volume = {17},
number = {11},
pages = {114057},
author = {Villanueva, D and Possner, A and Neubauer, D and Gasparini, B and Lohmann, U and Tesche, M},
title = {Mixed-phase regime cloud thinning could help restore sea ice},
journal = {Environmental Research Letters}}

@article{Korolev2022,
author = {Korolev, Alexei and Milbrandt, Jason},
title = {How Are Mixed-Phase Clouds Mixed?},
journal = {Geophysical Research Letters},
volume = {49},
number = {18},
pages = {e2022GL099578},
doi = {https://doi.org/10.1029/2022GL099578},
eprint = {https://agupubs.onlinelibrary.wiley.com/doi/pdf/10.1029/2022GL099578},
note = {e2022GL099578 2022GL099578},
year = {2022}
}

@misc{Wang2025JAMES,
	title = {Inverse {Mapping} of the {Collision} {Kernel} and {Wall} {Flux} {Scaling} in a {Tall} {Convection}-{Cloud} {Chamber} {Using} {Local} {Sensors} and {Knowledge}-{Informed} {Deep} {Learning}},
	doi = {10.22541/essoar.175745432.22699159/v1},
	urldate = {2025-11-06},
	publisher = {Preprints},
	author = {Wang, Aaron and Jiang, Peishi and Burrows, Susannah Marie and Glienke, Susanne and Ovchinnikov, Mikhail and Mahfouz, Naser},
	month = sep,
	year = {2025},
}

@article{Thomas2025JAMES,
	title = {Enhancing {Turbulent} {Mixing} and {Microphysical} {Uniformity} in a {Tall} {Convection}‐{Cloud} {Chamber} {Through} {Idealized} {Heterogeneity} of {Boundaries}},
	volume = {17},
	issn = {1942-2466, 1942-2466},
	doi = {10.1029/2025MS005316},
	language = {en},
	number = {11},
	urldate = {2025-11-06},
	journal = {Journal of Advances in Modeling Earth Systems},
	author = {Thomas, Lois and Yang, Fan and Wang, Aaron and Ovchinnikov, Mikhail and Pressel, Kyle G. and Shaw, Raymond A.},
	month = nov,
	year = {2025},
	pages = {e2025MS005316},
}

@manual{chen_2024_15626802,
  title        = {Case description for International Cloud Modeling
                   Workshop (ICMW) 2024 Mixed-Phase Cloud Chamber
                   Simulation Intercomparison
                  },
  author       = {Chen, Sisi and
                  Wang, Aaron and
                  Krueger, Steven},
  month        = may,
  year         = 2024,
  doi          = {10.5281/zenodo.15626802},
}

@article{Chandrakar2020JFM,
  title={Supersaturation fluctuations in moist turbulent Rayleigh--B{\'e}nard convection: A two-scalar transport problem},
  author={Chandrakar, Kamal Kant and Cantrell, Will and Krueger, Steven and Shaw, Raymond A and Wunsch, Scott},
  journal={Journal of Fluid Mechanics},
  volume={884},
  pages={A19},
  year={2020},
  publisher={Cambridge University Press}
}

@article{lanotte2009cloud,
  title={Cloud droplet growth by condensation in homogeneous isotropic turbulence},
  author={Lanotte, Alessandra S and Seminara, Agnese and Toschi, Federico},
  journal={Journal of the Atmospheric Sciences},
  volume={66},
  number={6},
  pages={1685--1697},
  year={2009}
}

@Article{niedermeier2020characterization,
author = {Niedermeier, D. and Voigtl\"ander, J. and Schmalfu{\ss}, S. and Busch, D. and Schumacher, J. and Shaw, R. A. and Stratmann, F.},
title = {Characterization and first results from LACIS-T: a moist-air wind tunnel to study aerosol--cloud--turbulence interactions},
journal = {Atmospheric Measurement Techniques},
volume = {13},
year = {2020},
number = {4},
pages = {2015--2033},
DOI = {10.5194/amt-13-2015-2020}
}

\end{document}